\documentclass[conference]{IEEEtran}
\usepackage{graphicx}
\usepackage{subfigure}
\usepackage{tablefootnote}
\usepackage{amsmath}
\usepackage{hyperref}
\usepackage{listings}
\usepackage{algorithm}
\usepackage{verbatim}
\usepackage{algpseudocode}
\usepackage{todonotes}
\usepackage{multirow}
\usepackage[nocompress]{cite}
\makeatletter
\g@addto@macro{\UrlBreaks}{\UrlOrds}
\makeatother

\definecolor{cgray}{gray}{0.90}
\definecolor{cgreen}{rgb}{0,0.6,0}
\definecolor{cmauve}{rgb}{0.58,0,0.82}
\newcommand{\para}[1]{\smallskip\noindent\textbf{\textit{#1}}}

\usepackage{hyperref}
    \definecolor{linkcolor}{rgb}{0.65,0,0}
    \definecolor{citecolor}{rgb}{0,0.4,0}
    \definecolor{urlcolor}{rgb}{0,0,0.65}
    \hypersetup{pdfpagemode=UseNone,pdfstartview=FitH,colorlinks=true, linkcolor=linkcolor, urlcolor=urlcolor, citecolor=citecolor}

\begin{document}

\newcommand{\detectorName}{FortuneTeller}
\newcommand{\detector}{\textit{\detectorName}}

\title{\detector: 
Predicting Microarchitectural Attacks \\ via Unsupervised Deep Learning
}	

\author{\IEEEauthorblockN{Berk Gulmezoglu\IEEEauthorrefmark{1},
		Ahmad Moghimi\IEEEauthorrefmark{1}, Thomas Eisenbarth\IEEEauthorrefmark{2} and
		Berk Sunar\IEEEauthorrefmark{1}}
	\IEEEauthorblockA{\IEEEauthorrefmark{1}Worcester Polytechnic Institute, \IEEEauthorrefmark{2}University of L\"ubeck}}
\maketitle

\begin{abstract}
    The growing security threat of microarchitectural attacks underlines the importance of robust security sensors and detection mechanisms at the hardware level. While there are studies on runtime detection of cache attacks, a generic model to consider the broad range of existing and future attacks is missing. Unfortunately, previous approaches only consider either a single attack variant, e.g. \textit{Prime+Probe}, or specific victim applications such as cryptographic implementations. Furthermore, the state-of-the art anomaly detection methods are based on coarse-grained statistical models, which are not successful to detect anomalies in a large-scale real world systems.
    
    Thanks to the memory capability of advanced Recurrent Neural Networks (RNNs) algorithms, both short and long term dependencies can be learned more accurately. Therefore, we propose \detector, which for the first time leverages the superiority of RNNs to learn complex execution patterns and detects unseen microarchitectural attacks in real world systems. \detector~models benign workload pattern from a microarchitectural standpoint in an unsupervised fashion, and then, it predicts how upcoming benign executions are supposed to behave. Potential attacks and malicious behaviors will be detected automatically, when there is a discrepancy between the predicted execution pattern and the runtime observation. 
    
    We implement \detector\ based on the available hardware performance counters on Intel processors and it is trained with 10 million samples obtained from benign applications. For the first time, the latest attacks such as Meltdown, Spectre, Rowhammer and Zombieload are detected with one trained model and without observing these attacks during the training. 
    We show that \detector~achieves the best false positive and false negative trade off compared to existing works under realistic workloads and target implementations with the highest F-score of $0.9970$. 
\end{abstract}

\section{Introduction}\label{sec:intro}

In the past decade, we have witnessed the evolution of microarchitectural side-channel attacks~\cite{yarom2014flush,irazoqui2014wait,evtyushkin2016jump,gruss2016prefetch,jang2016breaking}, from being considered as a nuisance and largely dismissed by chip manufacturers to becoming frontpage news. The severity of the threat was demonstrated by the Spectre~\cite{kocher2018spectre} and Meltdown~\cite{lipp2018meltdown} attacks, which allow a user with minimum access right to easily read arbitrary locations in the memory by exploiting the transient effect of illegal instruction sequences. This was followed by a plethora of attacks~\cite{maisuradze2018ret2spec,van2018foreshadow,schwarz2019zombieload,minkin2019fallout} either extending the scope of the microarchitectural flaws or identifying new leakage sources. It is noteworthy that these critical vulnerabilities managed to stay hidden for decades. Only after years of experimentation, researchers managed to gain sufficient insight into, for the most part, the unpublished aspects of these platforms. This leads to the point that they could formulate fairly simple but very subtle attacks to recover internal secrets. 
Therefore, the natural question becomes: how can we discover dormant vulnerabilities and protect against such subtle attacks? A fundamental approach is to eliminate the leakage altogether by using formal analysis. However, given the tremendous level of complexity of modern computing platforms and lack of public documentation, formal analysis of the hardware seems impractical in the near future. What remains is the modus operandi: {\bf leaks are patched as they are discovered by researchers through inspection and statistical analysis}. 

Countermeasures for microarchitectural side-channel attacks focus on the operating system (OS) hardening~\cite{LiuEtAl2016,gruss2017kaslr}, software synthesis~\cite{rane2015raccoon,cauligi2017fact} and analysis~\cite{almeida2013formal,weiser2018data,wichelmann2018microwalk}, and static~\cite{irazoqui2016mascat} or dynamic~\cite{Zhang2016cloudradar,chiappetta2016real,briongos2018cacheshield} detection of attacks. Static analysis is performed by evaluating the untrusted software against known malicious code patterns without running it on a target platform~\cite{irazoqui2016mascat}. Alternatively, dynamic analysis aims to detect malicious behaviors in the system by analyzing the runtime footprint of the running processes~\cite{chiappetta2016real}. Existing works on dynamic detection of microarchitectural attacks are based on collecting footprints from the hardware performance counters (HPCs) and limited modeling of malicious behaviors \cite{chiappetta2016real,mushtaq2018nights,Zhang2016cloudradar,briongos2018cacheshield}. A crucial challenge for both detection techniques is the shortage of knowledge about new attack vectors. Therefore, modeling malicious behaviors for undiscovered attacks and accurately distinguishing them from benign activities are open problems. Moreover, microarchitectural attacks are in infancy, and supervised learning models, which are used as attack classifier~\cite{mushtaq2018nights}, are not reliable to detect known attacks due to the insufficient amount and imprecise labeling of the data. Hence, unsupervised methods are more promising to adapt the detection models to real world scenarios.

Anomaly-based attack detection, which has been also studied in other security applications~\cite{shabtai2012andromaly,feizollah2013study}, aims to address the aforementioned challenge by only modeling the benign behaviors and detecting outliers. While there have been several efforts on anomaly-based detection of cache attacks~\cite{briongos2018cacheshield,chiappetta2016real}, modern microarchitectures have a diverse set of components that suffers from side-channel attacks~\cite{evtyushkin2016jump,yarom2017cachebleed,gruss2016prefetch,moghimi2018memjam}. Thus, detection techniques will not be practical and usable, if they do not cover a broad range of both known and unseen attacks. This requires more advanced learning algorithms to comprehensively model the entire behavior of the microarchitecture. On the other hand, statistical methods for anomaly detection are not sufficient to analyze millions of events that are collected from a very complex system like the modern microarchiecture. A major limitation of the classical statistical learning methods is that they use a hand-picked set of features, which wastes the valuable information to characterize the benign execution patterns. As a result, these techniques fail at building a generic model for real-world systems.

The latest advancements in Deep Learning, especially in Recurrent Neural Networks (RNNs), shows that time dependent tasks such as language modeling~\cite{sundermeyer2012lstm}, speech recognition~\cite{sak2014long} can be learned and upcoming sequences are predicted more efficiently by training millions of data samples. Similarly, computer programs are translated to processor instructions, and the corresponding microarchitectural events have time dependent behaviors. Modeling the sequential flow of these events for benign applications is extremely difficult by using logic and formal reasoning due to the complexity of the modern microarchitecture. We claim that these time dependent behaviors can be modeled in a large scale by observing sufficient number of benign execution flows. Since the long-term dependencies in the time domain can be learned with a high accuracy by training Long-short term memory (LSTM) and Gated Recurrent Unit (GRU) networks, the fingerprint of benign applications in a processor can also be learned efficiently. In addition, a challenging task of choosing the features of benign applications can be done automatically by LSTM/GRU networks in the training phase without any expert input.

\textbf{Our Contribution:}
We propose \detector\ which is the first generic detection model/technique for microarchitectural attacks. \detector~learns the benign behavior of hardware/software systems by observing microarchitectural events, and classifies any outlier that does not conform to the trained model as malicious behavior. \detector~can detect unseen microarchitectural attacks, since it only requires training over benign execution patterns. 

In summary, we propose \detector\ which:

\begin{itemize}
    \item is a generic detection technique, that can be applied to detect attacks on other microarchitectures and execution environments.
    \item for the first time, can detect various attacks automatically, disregarding the victim application, including cryptogrpahic implementations, browser passwords, secret data in kernel environment, bit flips and so on. 
    \item can detect attacks that were not observed during the training, or future attacks that may be introduced by the security community.
\end{itemize}

More specifically, we show: 

\begin{enumerate}
     \item different types of hardware performance counters can be used as the most optimum security sensor available on the commodity processors.
     \item how to capture the system-wide low-level microarchitectural traces and learn noisy time-dependent sequences through advanced RNN algorithms by training a more advanced and generic model. 
    \item \detector\ performs better by comparing it to the state-of-the art detection techniques. 
    \item we can detect malicious behavior dynamically in an unsupervised manner including stealthy cache attacks (Flush+Flush), transient execution attacks (Meltdown, Spectre, Zombieload) and Rowhammer.
\end{enumerate}

\subsection{Outline:}
The rest of the paper is organized as follows: \autoref{sec:background} provides background information on microarchitecural attacks, performance counters and RNNs. Then, \autoref{sec:related_work} gives an overview of previous works.  \autoref{sec:methodology} outlines the methodology and implementation of \detector. Also, information on our benign and attack dataset and performance counter selection are given. \autoref{sec:results} evaluates the results. The comparison with the prior works is given in~\autoref{sec:comparison}. Finally, \autoref{sec:discussion} discusses the results and \autoref{sec:conclusion} concludes our work.

\section{Background}\label{sec:background}

\subsection{Microarchitectural Attacks}

Modern computer architecture has a tremendously complex and optimized design. In order to improve the performance, several low-level features have been introduced such as speculative branches, out-of-order executions and shared last level cache (LLC). All these components are potential targets for microarchitectural attacks. Therefore, the following paragraphs give insight into microarchitectural attacks, which are examples of attacks that can be detected by \detector.

\para{Flush+Reload (F+R)} The LLC is shared among all cores in the processor. Flush+Reload attack~\cite{yarom2014flush} aims to track accesses to specific cache lines by using the \textit{clflush} instruction. First, adversary flushes the victim cache line. Then, the victim executes some instructions. Finally, the adversary reloads the same cache line and measures the access time. Flush+Reload attack is mostly used to recover cryptographic keys~\cite{yarom2014recovering}, which is applicable to perform attacks on systems with enabled memory deduplication such as cloud environments~\cite{IrazoquiEtAl2015, gulmezouglu2015faster}. 

\para{Flush+Flush (F+F)} Flush+Flush attack uses the \textit{clflush} instruction to flush the specific cache lines~\cite{GrussEtAl2016a}. Instead of measuring the time to access a cache line, the execution time of the \textit{clflush} instruction is measured. This method is considered as a stealthy attack against detection methods, since the number of introduced cache misses is low by this attack. Flush+Flush attack is used to exploit the T-table implementation of AES and user's keystrokes~\cite{GrussEtAl2016a}.

\para{Prime+Probe (P+P)} In a Prime+Probe attack, an adversary aims to fill an entire cache set, and then, measures the access time to the same cache set~\cite{tromer2010efficient}. If a victim evicts any of the adversary's cache line from the set, the adversary will observe access latency which leaks information about the victim's memory access pattern. While it has a lower resolution compared to Flush+Reload and Flush+Flush, it has a broader applicability. Prime+Probe attack was applied in the cloud environment to steal secret keys~\cite{ZhangEtAl2012, inci2016cache, irazoqui2015s}, Javascript to detect the visited webpages~\cite{OrenEtAl2015} and mobile phones to detect applications and user input~\cite{lipp2016armageddon, gulmezoglu2018undermining}.

\begin{figure*}[!t]
	\centering
	\includegraphics[width=0.85\textwidth]{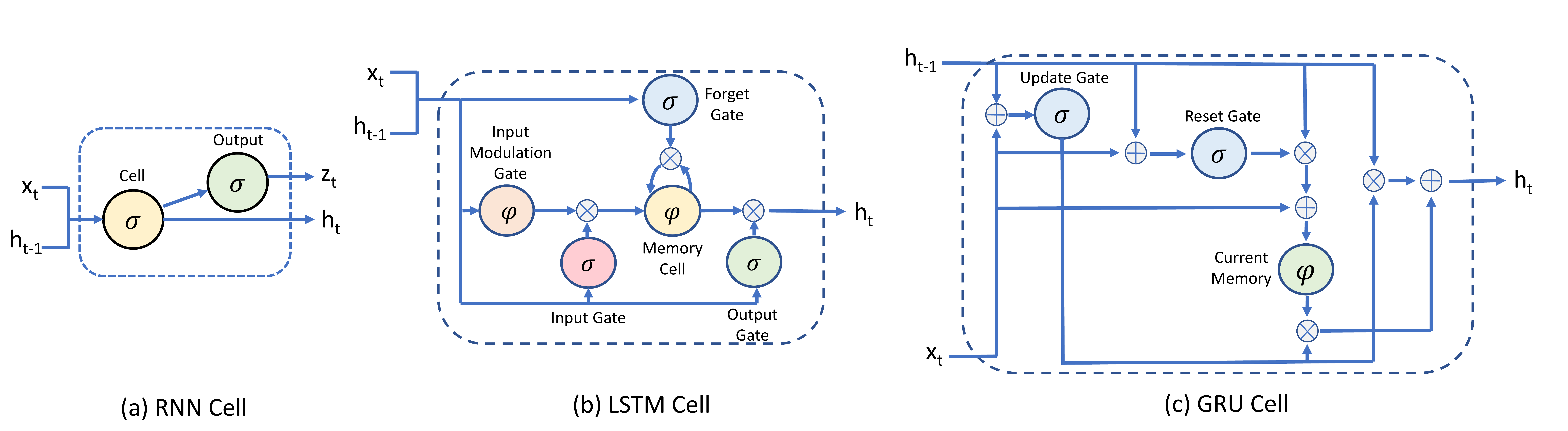}
	\caption{RNN, LSTM, GRU cells}
	\label{fig:RNN}
\end{figure*}
\subsection{Hardware Performance Counters (HPCs)}

\para{Rowhammer} DRAM cells have the possibility to leak charge over time. Rowhammer~\cite{gruss2016rowhammer} triggers the leak by accessing neighboring rows repeatedly. This leads to bit flips, which enables adversaries with low access right to gain system privileges~\cite{seaborn2015exploiting}. \textit{clflush} instruction is also commonly used to increase repeated access to the DRAM by bypassing the cache~\cite{aweke2016anvil}.

\para{Spectre} Spectre attacks exploit speculative branches~\cite{kocher2018spectre}. This attack is able to read memory addresses, which do not belong to the adversary by misusing the branch prediction. Therefore, sensitive data such as credentials stored in the browser can be leaked from the victim's memory space. Spectre is also effective against the SGX environment to compromise the trusted execution~\cite{van2018foreshadow,chen2018sgxpectre}. 

\para{Meltdown} Meltdown attack focuses on out-of-order execution to read kernel memory addresses~\cite{lipp2018meltdown}. The victim's secret, which is loaded into the registers will be mapped to different cache lines. Flush+Reload is used to determine if a specific cache line has been accessed. An adversary with only user privileges can perform this attack to read the content of kernel address space. The same concept has also been applied to Intel SGX~\cite{van2018foreshadow} to bypass the hardware supported memory isolation.

\para{ZombieLoad} Meltdown-style attacks can also specifically leak data from various microarchitectural resources such as store buffer~\cite{minkin2019fallout}, line fill buffer (LFB)~\cite{schwarz2019zombieload} and load ports~\cite{ridl}. ZombileLoad attack leaks data from memory load operations that are executed in other user processes and kernel context. Faulting/assisting loads that are executed by a malicious process can retrieve the stale data belonging to other security domains. This data that has been falsely forwarded from the shared resources may include secrets such as cryptographic keys or website URLs which can be transmitted over a covert channel such as the Flush+Reload technique.

\textit{Hardware Performance Counters (HPCs)} store low-level hardware-related events in the CPU. These events are tracked as counters that are available through special purpose registers. There are various performance events available in processors. The counters are used to collect information about the system behavior while an application is running. They have been used by researchers to reverse-engineer the internal design choices in the processor~\cite{maurice2015reverse}, or to increase the performance of the software by analyzing the bottlenecks~\cite{adhianto2010hpctoolkit}. These low-level counters are provided on all major architectures developed by ARM, Intel, AMD, and NVIDIA. 

There are various tools to program and read performance counters. Intel PCM~\cite{intelpcm} supports both core and offcore counters on Intel processors. The core counters give access to events within a single core of a processor, while the offcore counters profile events and activities across the cores and within the processor's die. This includes some of the events related to the integrated memory controller and the Intel QuickPath Interconnect which is shared by all cores. Before using the performance counters, we need to program each of them to monitor a specific event. Afterwards, the counter state can be sampled. In this work, we only focused on core counters, since the offcore counters have a small variety. 

\subsection{Recurrent Neural Networks (RNNs)}

RNNs are a type of Artificial Netural Network algorithm, which is used to learn and predict the sequential data. RNNs are mostly applied to speech recognition and currently used by Apple's Siri~\cite{AppleSiri} and Google's Voice Search~\cite{chiu2018state}. The reason behind the integration of RNNs into real world applications is that it is the first algorithm to remember the temporal relations in the input through its internal memory. Therefore, RNNs are mostly preferred for tasks where sequential data is involved.  

In a typical RNN structure, the information cycles through a loop. When the algorithm needs to make a decision, it uses the current input $x_t$ and hidden state $h_{t-1}$ where the learned features from the previous data samples are kept as shown in~\autoref{fig:RNN}a. Basically, a RNN algorithm produces output based on the previous data samples, and provides the output as a feedback into the network. However, traditional RNN algorithms are not good at learning the long-term sequences because the amount of extracted information converges to zero with the increasing time steps. In other words, the gradient is vanished and the model stops learning after long sequences. In order to overcome this problem, two algorithms were introduced, as described below: 

\subsubsection{Long-Short Term Memory}

Long-Short Term Memory (LSTM) networks are modified RNNs, which essentially extends the internal memory to learn longer time sequences. LSTM networks consist of memory cell, input, forget and output gates as shown in~\autoref{fig:RNN}b. The memory cell keeps the learned information from the previous sequences. While the cell state is modified by the forget gate, the output of the forget gate multiplies the specific positions in the input matrix by $0$ to forget and by $1$ to keep the information. The forget gate equation is as follow; $f_{t}=\sigma(W_f[h_{t-1},x_t]+b_t)$, where sigmoid function is applied to the weighted input and the previous hidden state. In the input gate, the useful input sections are determined to be fed into the cell state. The input gate equation is $i_t=\sigma(W_i[h_{t-1},x_t]+b_i)$, where sigmoid function is used as an activation function. This gate is combined with the input modulation gate to switch the cell state to forget memory. The activation function for input modulation gate is \textit{tanh}. Finally, the output gate passes the output to the next hidden state by applying the equation $o_t=\sigma(W_o[h_{t-1},x_t]+b_o)$, where \textit{tanh} is used as an activation function. Therefore, LSTM networks can select distinct features in the time sequence data more efficiently than RNNs, which enables learning the long-term temporal relations in the input.

\begin{table*}[t]
    \small
	\centering
	\caption{Comparison with prior works: \detector\ is able to detect attacks that were unseen during the training such as Flush+Flush (F+F), Prime+Probe (P+P), Flush+Reload (F+R), Spectre, Meltdown and Rowhammer. In contrast to prior works, \detector\ is fully unsupervised and is agnostic to the target application.}
	\begin{tabular}{| c | c | c | p{2,2cm} | p{3cm} | }
		\hline
		 Prior Works & Learning Algorithm & Approach & Detected Attacks & Target Implementations\\
		\hline
		Mushtaq et al.~\cite{mushtaq2018nights} & LDA/LR/SVM & Supervised & F+F, F+R & Crypto\\
		Zhang et al.~\cite{Zhang2016cloudradar} & DTW & Semi-Supervised & P+P, F+R & Crypto/Hash \\
		Chiappetta et al.~\cite{chiappetta2016real} & GS & Unsupervised & P+P, F+R & Crypto \\
		Briongos et al.~\cite{briongos2018cacheshield} & CPD & Unsupervised & F+F, F+R, P+P & Crypto\\
		\hline
		 & 	 &  & 	F+F, F+R, P+P & \\
		 & 	 &  & 	\textbf{Rowhammer} & \textbf{Benchmarks}\\ 
		\textbf{Our Work}& \textbf{LSTM/GRU} & Unsupervised & \textbf{Spectre Meltdown Zombieload} & \textbf{Real-world Apps}\\
		\hline
	\end{tabular}
	\label{table:comparison}
\end{table*}

\subsubsection{Gated Recurrent Unit}

Gated Recurrent Unit (GRU) is improved version of RNNs. GRU uses two gates called, update gate and reset gate. The update gate uses the following equation: $z_t=\sigma(W_zx_t+U_zh{t-1})$. Basically, both current input and the previous hidden state are multiplied with their own weights and added together. Then, a sigmoid activation function is applied to map the data between $0$ and $1$. The importance of the update gate is to determine the amount of the past information to be passed along to the future. Then, the reset gate is used to decide how much of the past information to forget. In order to calculate how much to forget, $r_t=\sigma(W_rx_t+U_rh_{t-1})$ equation is used, where the previous hidden state and current input are multiplied with their corresponding weights. Then, the results are summed and sigmoid function is applied. The output is passed to the current memory cell which stores the relevant information from the past. It is calculated as $h_{t}{'}=tanh(Wx_t+r_t \odot Uh_{t-1})$. The element-wise product between reset gate and weighted previous hidden layer state determines the information to be removed from previous time steps. Finally, the current information is calculated by the equation $h_t=z_t \odot h_{t-1}+(1-z_t) \odot h_t$. The purpose of this part is to use the information obtained from update gate and combine both reset and update gate information. Hence, while the relevant samples are learned by update gate, the redundant information such as noise is eliminated by reset gate.

In this work, the RNN algorithms are used in an unsupervised fashion where there is no need for separate validation dataset in the training phase. The validation error is calculated for each prediction in the next timestamp and the total validation error is given after each epoch.

\section{Related Work}\label{sec:related_work}

\subsection{Detecting Attacks using HPCs}

Low-level performance monitoring events such as HPCs have been used as security sensors to detect malicious activities~\cite{malone2011hardware,herath2015these}. Similar to~\cite{yuan2011security,xia2012cfimon}, \emph{Numchecker}
\cite{wang2013numchecker} and \emph{Confirm}
\cite{wang2015confirm} adopt these sensors to detect 
control flow violations, which are applied to \textit{rootkits} and \textit{firmware modifications}, respectively. In addition, classical ML algorithms such as support vector machines (SVMs) and k-nearest neighbors (KNNs) are adapted to naive heuristic-based techniques for multi-class classification
\cite{bahador2014hpcmalhunter, demme2013feasibility}. The latter explores neural network in a supervised fashion~\cite{demme2013feasibility}. Tang et al.~\cite{TangEtAl2014} train One-Class Support Vector Machine (OC-SVM) with benign system behavior and detect the malware in the system. 

Despite the detection of malware and rootkits in the system, HPCs have also been used to detect microarchitectural attacks. Since our work focuses on microarchitectural attack detection, the features of prior approaches and our detection technique are compared in~\autoref{table:comparison}. Firstly, Chiappetta et al.~\cite{chiappetta2016real} proposes to monitor HPCs and the data is analyzed by using Gaussian Sampling (GS) or probability density function (pdf) to detect the anomalies on cryptographic implementations dynamically. Later, Zhang et al.~\cite{Zhang2016cloudradar} apply Dynamic Time Wrapping (DTW) to catch cryptographic implementation executions in the victim VMs. Then, the number of cache misses and hits in the attacker VMs are monitored during the execution of the sensitive operations. Briongos et al.~\cite{briongos2018cacheshield} implement Change Point Detection (CPD) technique to determine the sudden changes in the time series data to detect F+F, F+R and P+P attacks. Finally, Mushtaq et al.~\cite{mushtaq2018nights} detect the cache oriented microarchitectural attacks with supervised Linear Discriminant Analysis (LDA), Support Vector Machine (SVM) and Linear Regression (LR) technique under various system loads. We further compare the most related works with \detector~in~\autoref{sec:comparison}.

\subsection{RNN Applications in Security}

RNN algorithms are applied to other security domains to increase the efficiency of defensive technologies. For instance, Shin et al.~\cite{shin2015recognizing} leverage RNNs to identify functions in the program binary. Once the model is trained with these function, the technique classifies the bytes to decide on whether it is the beginning of the function or not. Similarly, Pascanu et al.~\cite{pascanu2015malware} apply RNNs to detect malware by training the APIs in an unsupervised way. The technique improves the true positive rate by 98\% compared to previous studies. In another study, Melicher et al.~\cite{melicher2016fast} introduce RNN-based technique to improve guessing attacks on password's resistance. This study shows better accuracy than Markov models. Furthermore, Du et al.~\cite{du2017deeplog} implement LSTM based anomaly detection to detect anomalies in the system. The LSTM model is trained with log data obtained from normal execution. Their results show that the traditional data mining techniques underperform LSTM model to detect the anomalies. Finally, Shen et al.~\cite{shen2018tiresias} apply LSTM and GRU networks to predict the next security events with a precision of up to 0.93. These studies indicate that RNN based security applications are commonly used in other challenging environments. 

\section{\detector}\label{sec:methodology}

\begin{figure*}[!t]
	\centering
	\includegraphics[width=.85\textwidth]{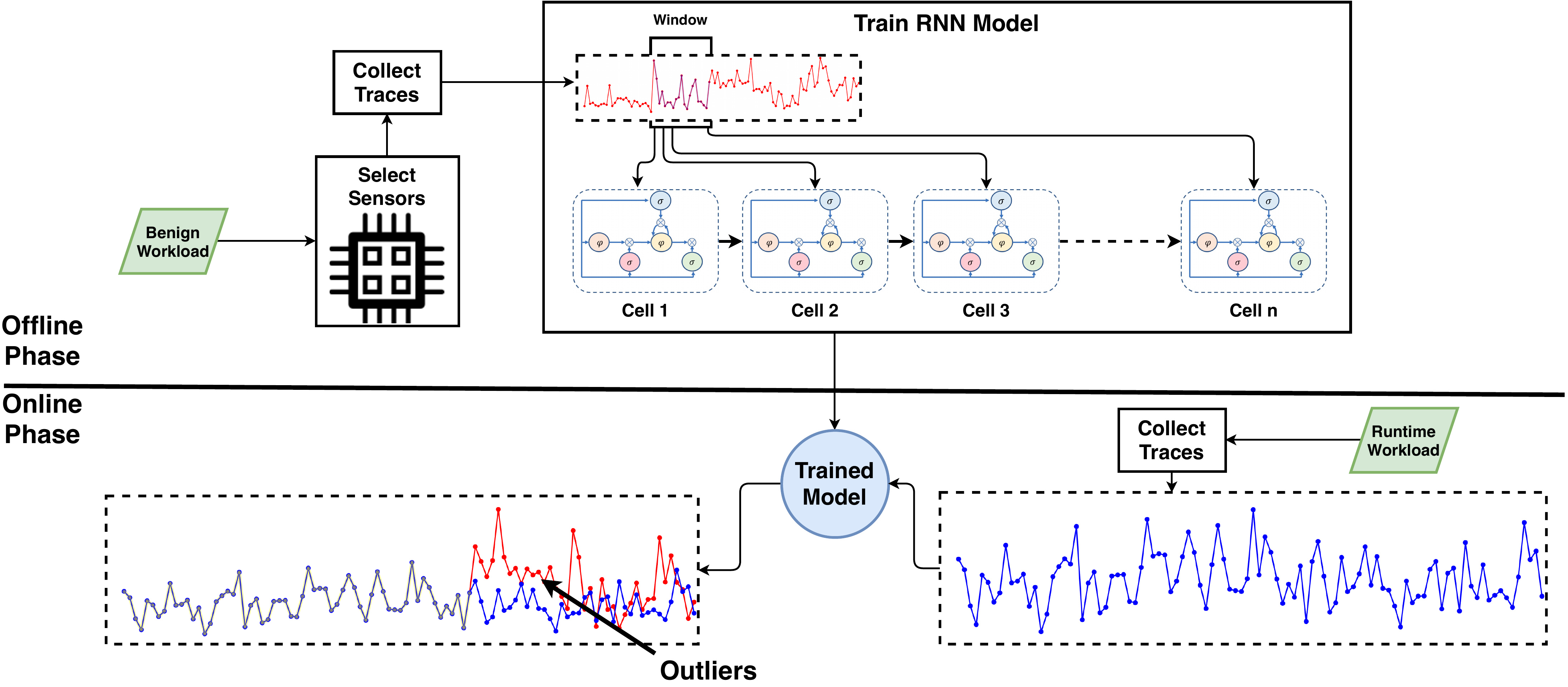}
	\caption{\detector implementation}
	\label{fig:method}
\end{figure*}

\subsection{Methodology}

Our conceptual design for \detector~consists of two phases as shown in~\autoref{fig:method}: In the offline phase, \detector~collects time sequence data from diverse set of benign applications by monitoring security sensors in the system. The collected data is used as the training data and it is fed into the RNN algorithm with a sliding window technique. The weights of the trained model are optimized by the algorithm itself since each data sample is also used as the validation. When there is no further improvement in the validation error, the training process stops. Once the RNN model is trained, it is ready to be used in a real time system. 

In the online phase, the real-time sequences are captured from the same security sensors and given as input to RNN models. The prediction of the next measurement for each sensor is made by the pre-trained RNN models, dynamically. If the mean squared error (MSE) between the predicted value and the real time sensor measurement is consistently higher than a threshold, the anomaly flag is set. The online phase is the actual evaluation of \detector~in a real world system.

Two separate detection models are trained with LSTM and GRU networks since they are known for their extraordinary capabilities in learning the long time sequences. Our purpose is to train an RNN-based detection model, which can predict the microarchitectural events in the next time steps with the minimal error. In our detection scenario, we consider a time series $X=\{\textbf{x}^{(1)},\textbf{x}^{(2)},\dots,\textbf{x}^{(n)}\}$, where each measurement $\textbf{x}^{(t)} \in R^m$ is an m-dimensional vector $\{x_{1}^{(t)},x_{2}^{(t)},\dots,x_{m}^{(t)}\}$ and each element corresponds to a sensor value at time $t$. As all temporal relations can not be discovered from millions of samples, a sliding window with a size of $W$ is used to partition the data into small chunks. Thereby, the input to RNN algorithm at time step $t$ is $\{x_{1}^{(t-W+1)},x_{2}^{(t-W+2)},\dots,x_{m}^{(t)}\}$, where the output is $\textbf{y}^{(t)}=\textbf{x}^{(t+1)}$. Note that, even though there is a fixed length sliding window in the problem formulation, the overall input size is not fixed. Finally, the trained model is saved to be used in real-world system.

To evaluate the trained model, test dataset is collected from benign applications and attack executions. The test dataset has the same structure with the training data, and is fed into the model to calculate the prediction error in the next time steps. The error at time step $t+1$ is $e^{(t+1)}$ which is equal to $1/m
\sum_{i=1}^{m}(y_{i}^{t+1}-x_{i}^{t+1})^2$. The model predicts the value of next measurement and then, the error for  is summed up to one value. 

To detect the anomalies in the system, a decision window $D$ and an anomaly threshold $\tau_{A}$ are used. If all the predictions in $D$ are higher than $\tau_{A}$, then the anomaly flag $F_A$ is set in the system in~\autoref{eq:anomaly}.

\begin{equation}
F_A= 
\begin{cases}
1, & \forall e^{(t+1)} \in D \geq\tau_{A}\\
0, & otherwise
\end{cases}
\label{eq:anomaly}
\end{equation}

The choice of $D$ directly determines the anomaly detection time. If $D$ is chosen as a small value, the attacks are discovered with a very small leakage. On the other hand, the false alarm risk increases in parallel, which is controlled by adjusting $\tau_{A}$. This trade-off is discussed further in~\autoref{sec:results}. 

\subsection{Implementation}

\subsubsection{Profiled Benchmarks and Attacks}\label{sec:dataset}

The main purpose of \detector~is to train a generic model by profiling a diverse set of benign applications. Therefore, selecting benign applications is utmost importance. For the benign application dataset, benchmark tests in Phoronix benchmark suite~\cite{phoronix} are profiled since the suite includes different type of applications such as cryptographic implementations, floating point and compression tests, web-server workloads etc. The complete list is given in Appendix,~\autoref{table:tests}. It is important to note that some benchmark tests have multiple sub-tests and all the sub-tests are included in both training and test phases. In addition to CPU benchmarks, we evaluate our detection models against system, disk and memory test benchmarks. In order to increase the diversity, the daily applications such web browsing, video rendering, Apache server, MySQL database and Office applications with several parameters are profiled for real-world examples. 

A subset of benign execution data is used to train our RNN models and then, the whole benign dataset is used to calculate the FPR (False Positive Rate) and TNR (True Negative Rate) of the models. In our work, FPs represents the benign applications which are classified as an attack/anomaly by the RNN model. If the benign application does not raise the alarm flag, it is considered as TNs. 

For the attack executions we include traditional cache attack techniques such as F+F, F+R and P+P attacks. Different from previous works, these attacks are applied on arbitrary memory blocks to avoid any assumption on the target implementation. Spectre (v1 and v2) and Meltdown are implemented to read secrets such as passwords in a pre-determined memory location. In addition, two types of Rowhammer attacks namely, one-sided and double-sided, are applied to have bit flips. In order to test the efficiency of \detector~we implemented a recent microarchitectural attack, Zombieload, to steal data across processes. For this purpose, a victim thread leaks pre-determined ASCII characters and the attacker reads the line-fill buffer to recover the secret. If the alarm flag is set during the execution of the attack, it is True Positive (TP). On the other hand, the undetected attack execution is represented by False Negative (FN).

\subsubsection{Performance Counter Selection}\label{sec:counters}

In our detection model, we leverage HPCs as security sensors. Although the number of available counters in a processor is more than 100, it is not feasible to monitor all counters concurrently. In an ideal system, we should be able to collect data from a diverse set of events to be able to train a generic model. However, due to the limited number of concurrently monitored events, we choose the most optimum subset of counters that give us information about common attacks. For this purpose, we perform a study of the best subset of performance counters. 

In our experiments, we leverage Intel PCM tool~\cite{intelpcm} to capture the system-wide traces. The set of counters in our experiments is chosen from \textit{core} counters. The main reason to choose core counters is the high variety of the available counters such as branches, cache, TLB, etc. The number of core counters tested in the selection method is 36. The complete list is given in Appendix,~\autoref{table:counters}. 

In the data collection step, a subset of the counters is profiled concurrently, since the number of counters monitored in parallel is limited to four in Intel processors. For each subset, a separate dataset collected until all 36 counters are covered. The training data is collected from 30 different Phoronix benchmark tests~\cite{phoronix} (1-30 in the~\autoref{table:tests}). In order to decide on the most suitable counters to detect microarchitectural anomalies, we collect a test dataset from 20 benchmark tests (1-56 in the~\autoref{table:tests}) and 6 microarchitectural attacks (174-179 in the~\autoref{table:tests}). The Zombieload attack is not included in the performance counter selection phase, since it was not released at that time. The sampling rate is chosen as 1 ms to have the minimal overhead in the system.

For every subset of counters one LSTM model is trained with a window size of $W=100$. The four dimensional data is given as an input to LSTM model and then, the final counters are selected based on their F-score given in Appendix,~\autoref{table:counters}. It is observed that some counters have better accuracy than other counters for specific attacks. For instance, branch related counters have high correlation for Meltdown and Spectre attacks. However, the F-score is also around 0.3 because real-world applications also use the branches heavily. One of the important outcome of selection phase is that speculative branches are commonly integrated in the benign applications. Therefore, the counter selection shows that branch counters are not useful to detect speculative execution attacks. Thanks to our LSTM based counter selection technique, finding the most valuable counters is fully automated and the success rate of detecting anomalies with low FPR and FNR is increased significantly. 

\begin{figure}[!t]
	\centering
	\subfigure[Instruction Cache Miss]{\includegraphics[width=0.47\textwidth]{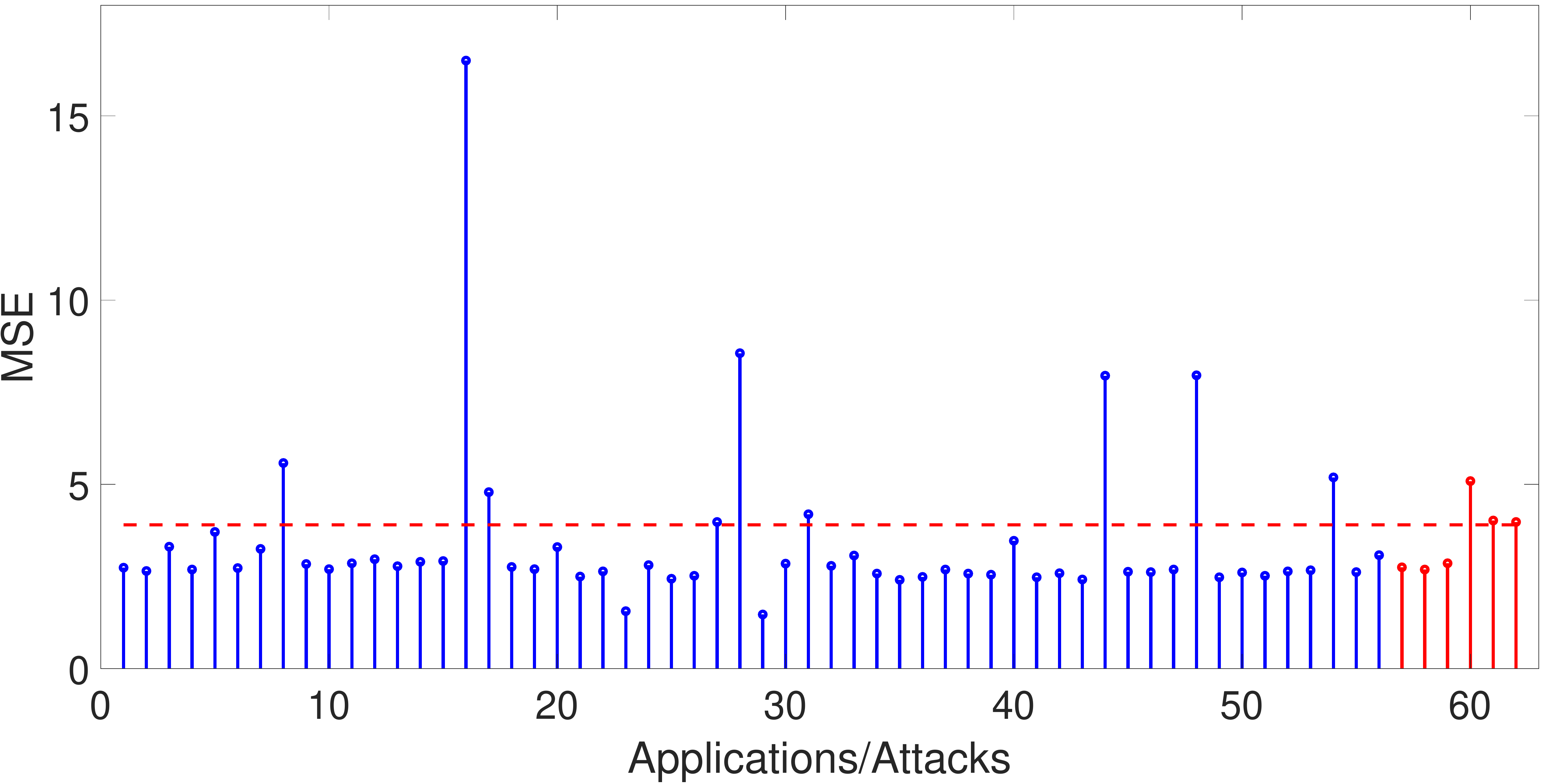}\label{fig:icache_miss}}
	\subfigure[Instruction Cache Hit]{\includegraphics[width=0.47\textwidth]{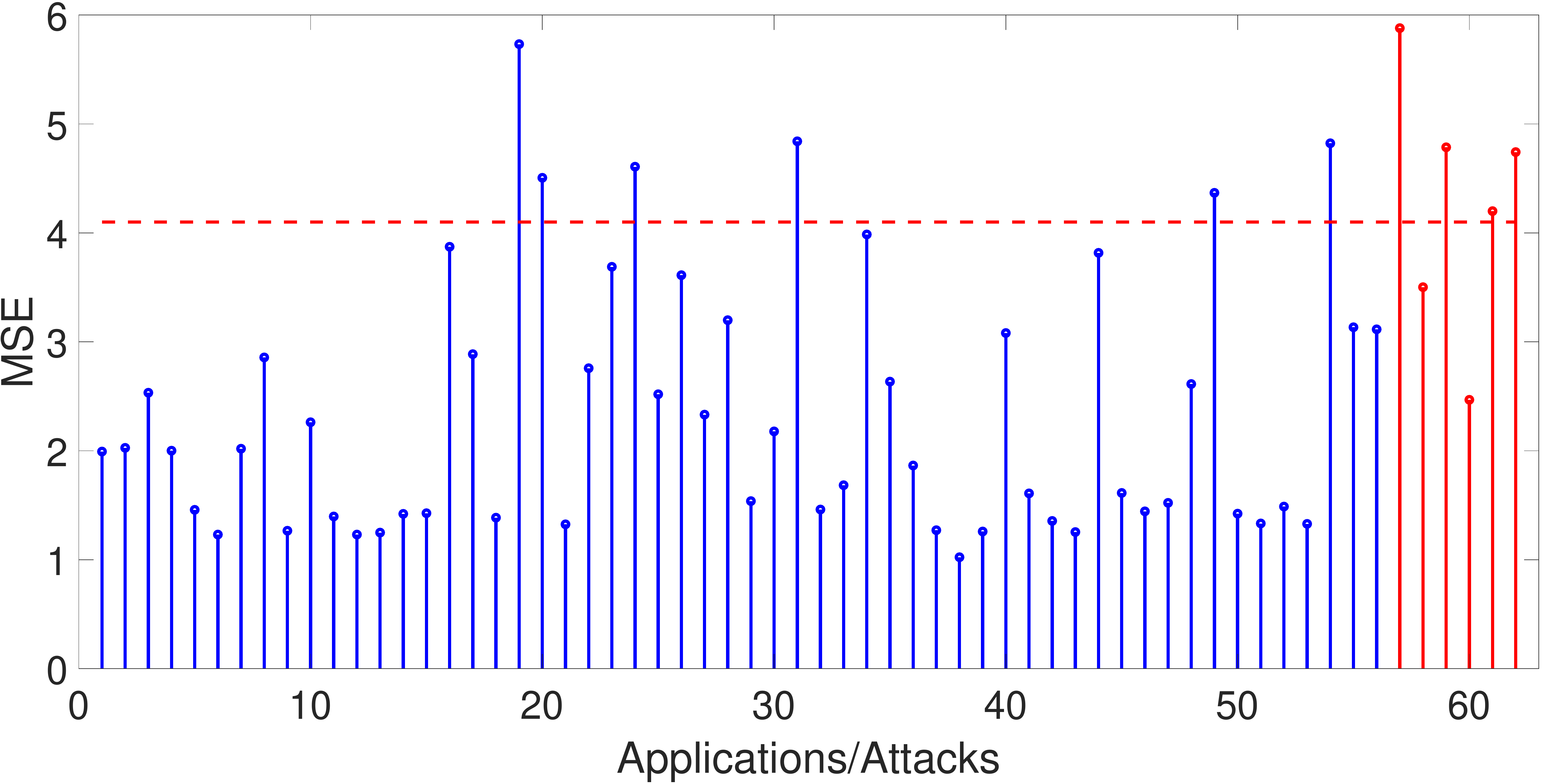}\label{fig:icache_hit}}
	\subfigure[LLC Miss]{\includegraphics[width=0.47\textwidth]{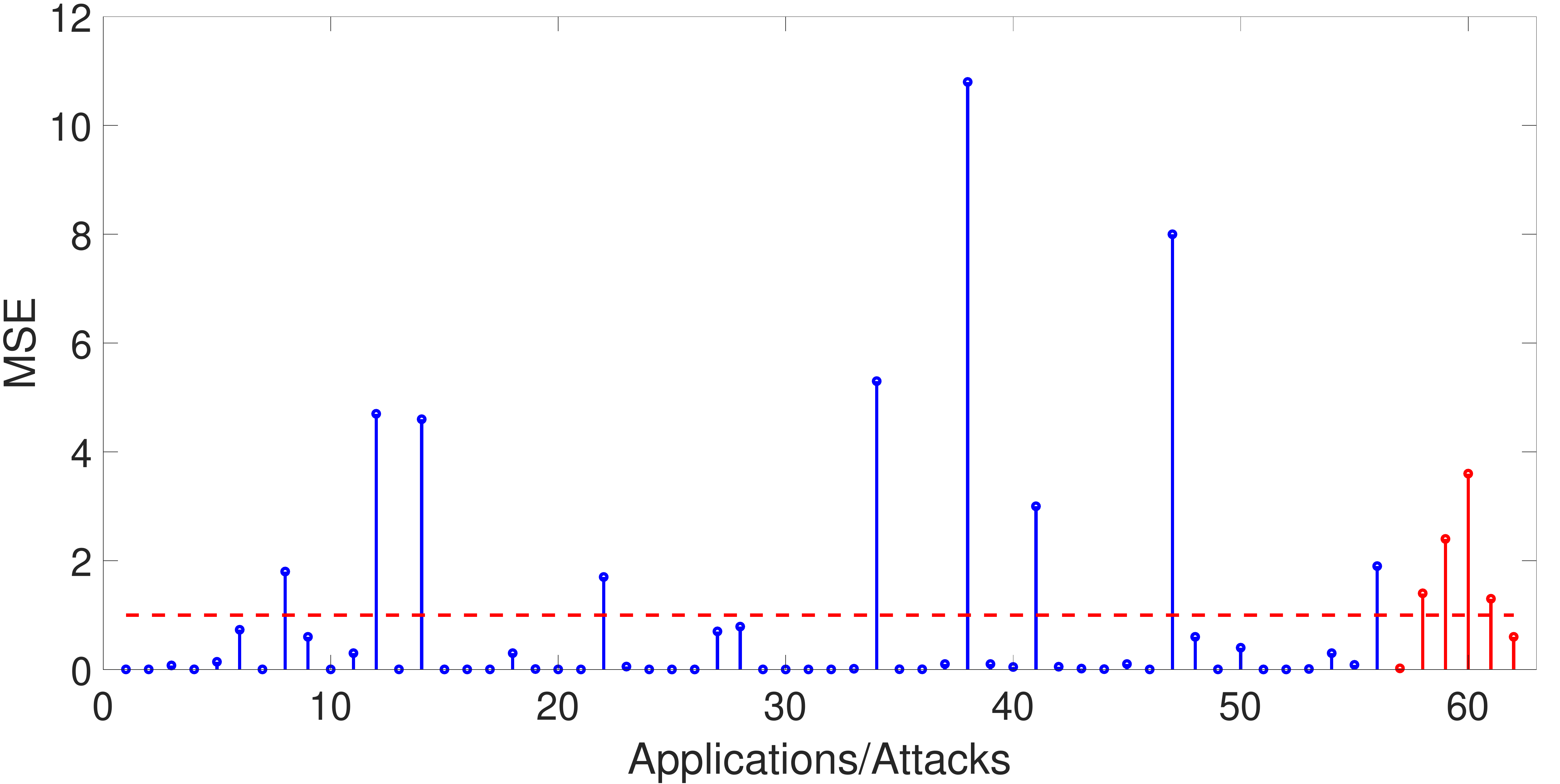}\label{fig:llc_miss}}
	\caption{Mean squared error rates for benign (blue) and attack (red) executions by monitoring (a) \textit{instruction cache miss}, (b) \textit{instruction cache hit}, and (c) \textit{LLC miss} counters. }
	\label{fig:counters}
\end{figure}

Even though it is allowed to choose up to 4 counters on the Intel server systems like Xeon, we selected 3 counters to profile for anomaly detection. The reason behind this is that in the desktop processors (Intel Core i5, i7) the programmable counters are limited to 3. The first selected counter is $L1\_Inst\_Miss$, which is more successful to detect Rowhammer, Spectre and Meltdown attacks with 14\% FPR, where the F-score is 0.7979. As a second counter, $L1\_Inst\_Hit$ is chosen, since Flush+Flush and Flush+Reload attacks are detected with a high accuracy and the F-score is 0.8137. The reason behind the high F-score is that the \textit{flush} instruction is heavily used in those attacks and the instruction cache usage also increases in parallel. Interestingly, Flush+Flush attack is known as a stealthy microarchitectural attack however, it is possible to detect it by monitoring instruction cache hit counter. The last selected counter is $LLC\_Miss$, which is successful to detect Rowhammer and Prime+Probe attacks with a high accuracy. These attacks cause frequent cache evictions in the LLC, which increases the number of anomalies in the $LLC\_Miss$ counter. These results show that the individual counters are not efficient to detect all the microarchitectural attacks. Therefore, there is a need for the integration of the aforementioned 3 counters to detect all the attacks with a high confidence rate.


\section{Evaluation}\label{sec:results}

In this section, we explain the experiments which are conducted to evaluate \detector. The experiments aim to answer the following research questions: 1) How does \detector~perform in predicting the next performance counter values for benign applications with the increasing number of measurements(\autoref{sec:RNN_training})? What is the lowest possible FPR for server (\autoref{sec:server}) and laptop environments (\autoref{sec:laptop})? 3) How does the size of sliding window affect the performance of \detector~(\autoref{sec:window_size})? 4) How realistic is real time protection with \detector~(\autoref{sec:dynamic})? 5) How much performance overhead is caused by \detector~(\autoref{sec:performance_overhead})?

\subsection{Experiment Setup}

\detector~is tested on two separate systems. The first system runs on an Intel Xeon E5-2640v3, which is a common processor used on server machines. It has 8 cores with 2.6 GHz base frequency and 20 MB LLC. The second device is used to illustrate a typical laptop/desktop machine, which is based on Intel(R) Core(TM) i7-8650U CPU with 1.90 GHz frequency. It has 8MB LLC and 2 cores in total.

Two types of RNN model namely, LSTM and GRU, are used to train \detector. The sliding window size, batch size and number of hidden LSTM/GRU layers are kept same in the training phase. Training of RNN models is done using the custom Keras~\cite{chollet2015keras} scripts together with the Tensorflow~\cite{abadi2016tensorflow} and GPU backend. The models are trained on a workstation with two Nvidia 1080Ti (Pascal) GPUs, a 20-core Intel i7-7900X CPU, and 64 GB of RAM. 

\subsection{RNN Model Training}\label{sec:RNN_training}

\begin{figure}[!t]
	\centering
	\includegraphics[width=0.48\textwidth]{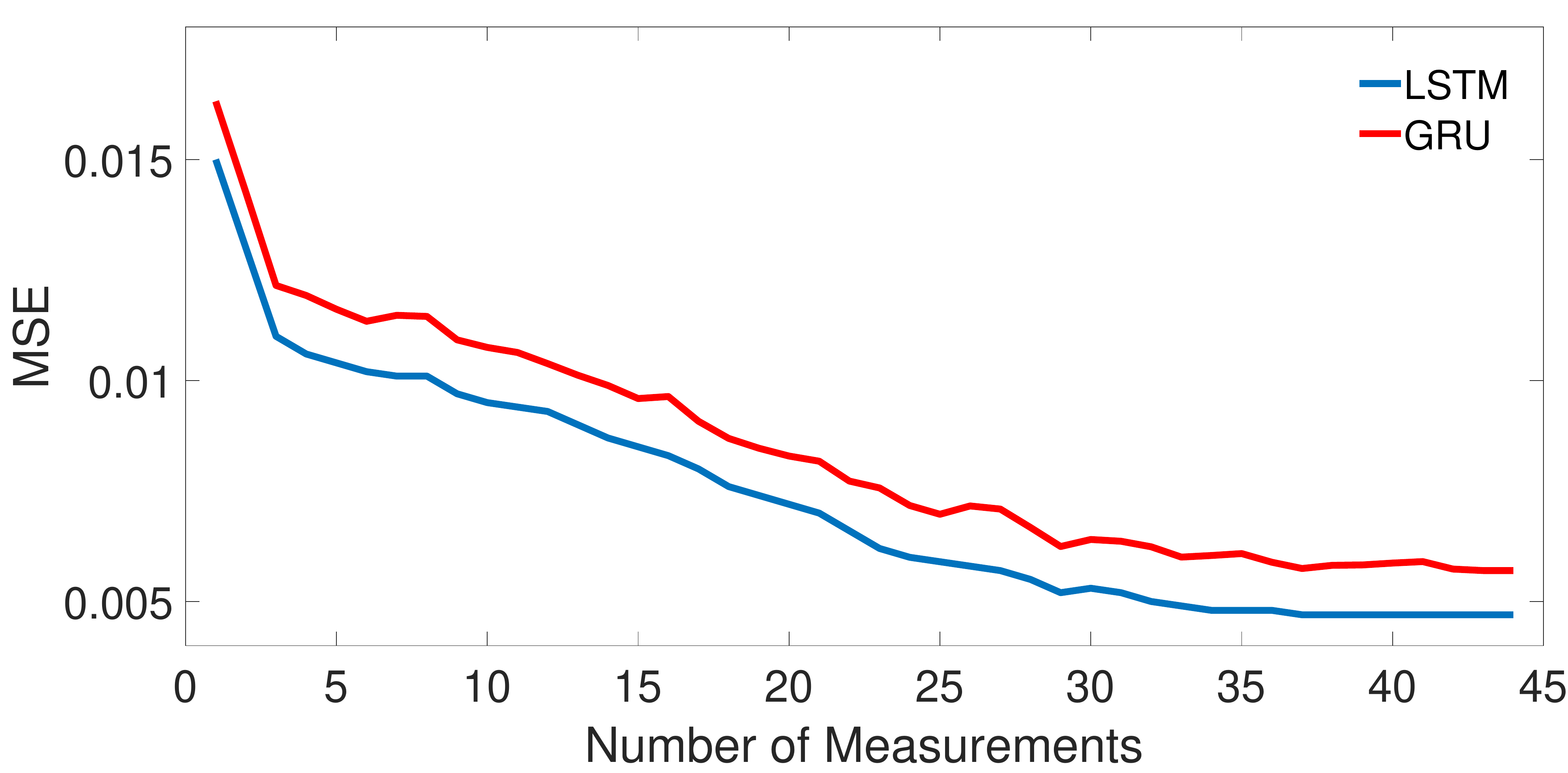}
	\caption{Validation error with increasing number of measurements for Gnupg benchmark}
	\label{fig:gnupg}
\end{figure}

To detect the anomalies in the system, the first step is to learn the pattern of the benign applications. This is not an easy task, since the chosen benchmarks and real-world applications have complicated fingerprint in the microarchitectural level with the system noise. Moreover, the fingerprint at each execution is not identical and the execution of the application takes several seconds, which makes it difficult to learn long-term relations in the data. Therefore, the required number of measurements from each individual application plays a crucial role to train the \detector~successfully. For this purpose, we choose 10 random benchmarks and a separate model for each of them is trained. The validation error obtained as a result of training is the critical metric to determine the capacity of the RNN algorithms as it indicates how well \detector~guesses the next counter value. The first RNN model is trained with only 1 measurement and the number of measurements is increased gradually up to 44. It is observed that there is no further improvement in the validation error after 36 measurements for both LSTM and GRU networks in~\autoref{fig:gnupg}. Note that, the training data is scaled to [0 1] and the validation error is the average error of the 3 counters.

\begin{figure}[!t]
	\centering
	\includegraphics[width=0.48\textwidth]{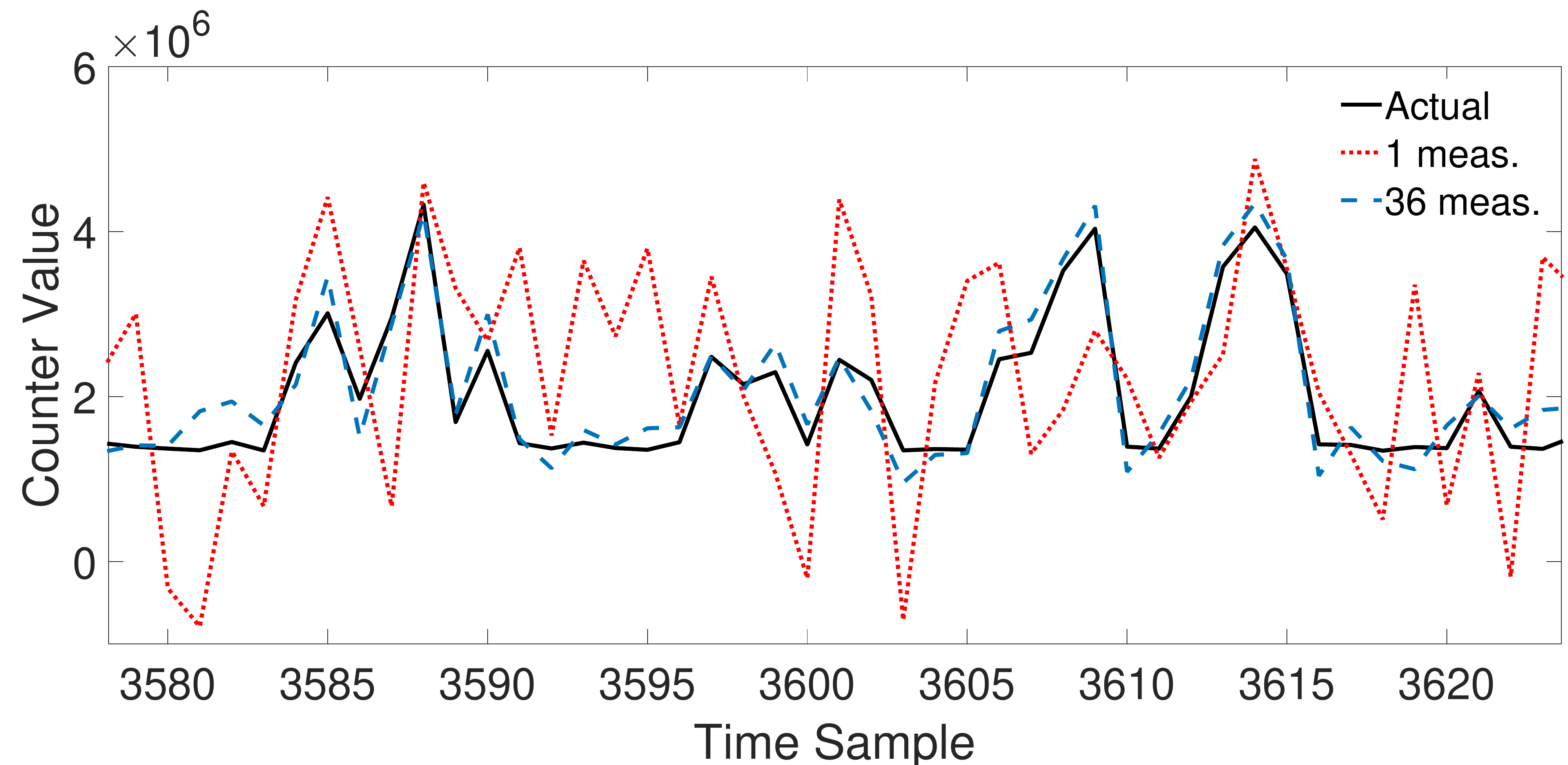}
	\caption{Prediction error in Gnupg for LSTM algorithm}
	\label{fig:gnupg_prediction}
\end{figure}

In~\autoref{fig:gnupg_prediction}, the prediction of $ICache.Hit$ counter value by using LSTM network is shown. The solid line represents the actual counter value while two other lines show the prediction values. When there is only one measurement to train the LSTM network, the prediction error is much higher. It means the trained model could not optimize the weights with small amount of data. On the other hand, once the number of measurements is increased to 36, the predictions are more consistent and close to actual counter value. The number of measurements directly affects the training time of the model. If the dataset is unnecessarily huge, the training time increases in linear. Therefore, it is decided to collect 36 measurements from each application in the training phase to achieve the best outcome from RNN algorithms in the real systems. With the accurate modeling of the benign behavior, the number of false alarms is reduced significantly. This is the main advantage of \detector, since the previous detection mechanisms apply a simple threshold technique to detect the anomalies when a counter value exceeds the threshold. In contrast, \detector~can predict the sudden increases in the counter values and the correct classification can be made more efficiently than before.

\subsection{Server Environment}\label{sec:server}

The first set of experiments is conducted in the server machine. Three core counters, $ICache.Miss$, $ICache.Hit$ and $LLC.Miss$ are monitored concurrently in the data collection process. The training dataset is collected with a sampling rate of 1 ms from $m=3$ core counters during the execution of benign applications. The dataset has 10 million samples (10,000 seconds) in total, collected from 67 randomly selected benchmark tests, 100 websites rendering in Google Chrome, Apache server/client benchmark, MySQL database and Office applications as listed in Appendix,~\autoref{table:tests}. Note that the idle time frames between the executions are excluded from dataset to avoid the redundant information in the training phase. Firstly, LSTM model is trained with the collected dataset where the input size is $3\times10,000,000$. The sliding window size is selected as $W=100$, which means the total number of LSTM units equals to 100. The further details of window size analysis is given in~\autoref{sec:window_size}. The training is stopped after 10 epochs since the validation error does not improve further. The validation error decreases to 0.0015. The training time for 10,000,000 samples takes approximately 4 days.

After the LSTM model is trained, a new dataset for the test phase is collected from counters by profiling 173 benign benchmark tests, 100 random websites, MySQL, Apache, Office applications and micro-architectural attacks. The length of the test data for each application would change, since our anomaly-based model has no assumption on the input length. Hence, the number of samples obtained from each application changes between 1000-20000. The number of samples for websites is around 1000, since the rendering process is extremely fast. However, some benchmarks have a longer execution time, which requires to collect data for a longer time. The remaining applications (Office, MySQL, Apache) are profiled for around 5 seconds.

Each application is monitored 50 times, and then, the test data is fed into the LSTM model to predict the counter values at the next time steps. Moreover, in order to make the test phase more realistic, the number of applications running concurrently is increased up to 5. The applications are chosen randomly from the test list in Appendix~\autoref{table:tests}, and started at the same time. 100 measurements are collected from concurrently running applications. In total, 25,000,000 samples are collected for the test phase.

\begin{figure}[!t]
	\centering
	\includegraphics[width=0.48\textwidth]{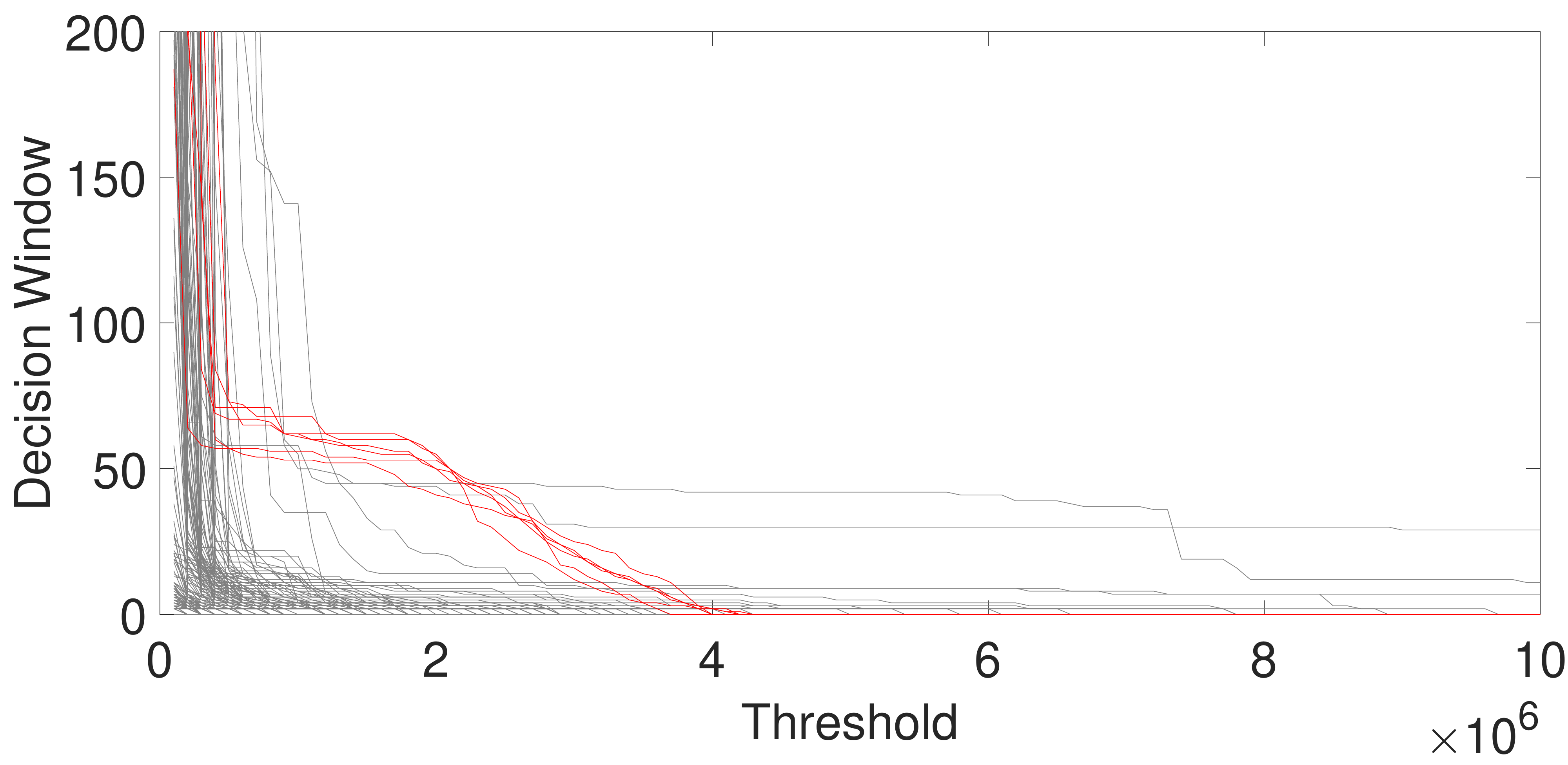}
	\caption{Threshold vs. Decision Window for benign applications (gray) and attack executions (red) using LSTM model in server scenario}
	\label{fig:server_lstm}
\end{figure}

The prediction is made for all three counters at each time step (every 1 ms), and then, the mean squared error $e^{(t+1)}$ is computed between the actual and predicted counter values. $e^{(t+1)}$ in the prediction step is used to choose optimal decision window $D$ and $\tau_{A}$ to detect the anomalies in the system. If the prediction error is higher than the threshold $\tau_{A}$ for $D$ samples, the application is classified as an anomaly. The threshold and decision window are chosen as to equalize FNR and FPR. The trend between $\tau_{A}$ for $D$ is given in~\autoref{fig:server_lstm}. For the lower $\tau_{A}$ values, the decision window is not applicable to detect the anomalies, since the benign applications and attack executions have higher error rates. Once the $\tau_{A}$ reaches $1.8 \times 10^6$, most of the attack executions are detected in $D=50$ samples. In other words, the microarchitectural attacks are caught in 50 ms by \detector. With the increasing $\tau_{A}$ values, the number of true positives begins decreasing, which yields to low detection rate.

The results show that P+P attack is the most difficult attack to be detected by the LSTM model in the server. This result is expected since P+P attack mostly focuses on specific cache sets and the cache miss ratio is smaller than other type of attacks. In addition, instruction cache is not heavily used by P+P attack, which makes the detection more difficult for \detector due to the lack of specific pattern. On the other hand, the highest TPR is obtained for Flush+Reload and Rowhammer attacks with 100\% and 0\% FNR. As these attacks increase the number of data cache misses and instruction hits through the extensive use of \textit{clflush} instruction, the fluctuation in the counter values is higher than the other types of attacks and benign applications. The accuracy of the predicting the next values decreases when the variance is high in the counters, thus, the prediction error increases in parallel. Since the higher prediction rate is a strong indicator of the attack executions in the system, \detector~detects them with a high accuracy. Note that, Zombieload is also detectable by the \detector, even though it was not included in the performance counter selection phase. This shows that \detector~can detect the unseen microarchitectural attacks with the current trained models.

\begin{figure}[!t]
	\centering
	\includegraphics[width=0.297\textwidth]{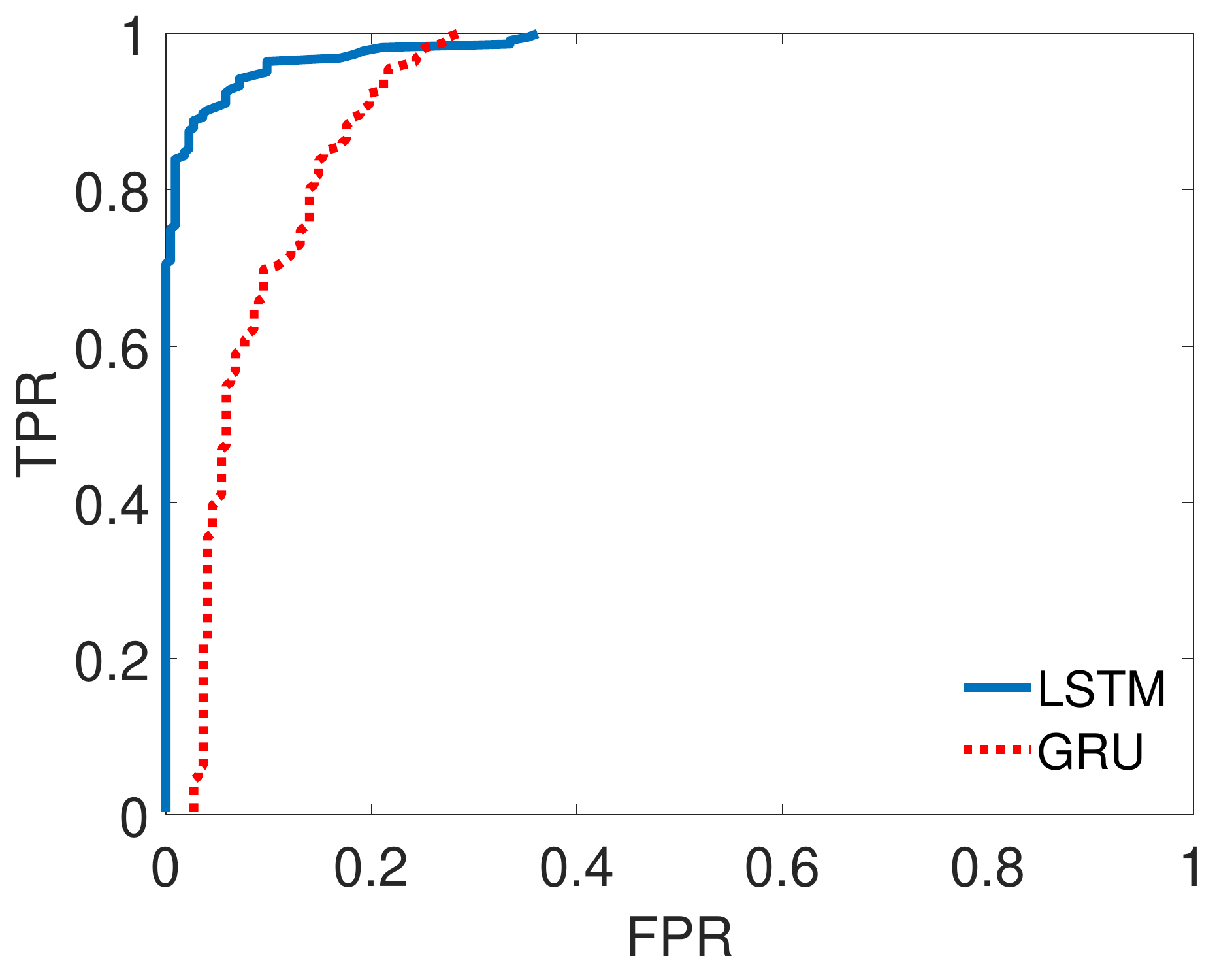}
	\caption{ROC curve for LSTM and GRU models in server scenario}
	\label{fig:server_ROC}
\end{figure}

The ROC curves in~\autoref{fig:server_ROC} indicate that LSTM networks have a better capability than GRU networks to detect the anomalies. The counter values are predicted with a higher error rate in GRU networks, which makes the anomaly detection harder. Some benign applications are always detected as anomaly by GRU, thus, the FPR is always high for different threshold values. The AUC (Area Under the Curve) for LSTM model is very close to perfect classifier with a value of 0.9840. On the other hand, the AUC for GRU model is 0.9125, which is significantly worse than LSTM model. There are several reasons behind the poor performance of GRU networks. The first reason is that GRU networks are not successful to learn the patterns of Apache server applications since there is a high fluctuation in the counter values. In addition, when the number of concurrently running applications increases, the false alarms increase drastically. On the other hand, LSTM networks are good at predicting the combination of patterns in the system. Therefore, the FPR is very small for LSTM model.

\subsection{Laptop Environment}\label{sec:laptop}

The experiments are repeated for the laptop environment to evaluate the usage of \detector. LSTM and GRU models are trained with 10 million samples, which is collected from benign applications. Since the laptops are mostly used for daily works, the counter values are relatively smaller than the server scenario. However, the applications stress the system more than the server scenario since the number of cores is lower. 

\begin{figure}[!t]
	\centering
	\includegraphics[width=0.48\textwidth]{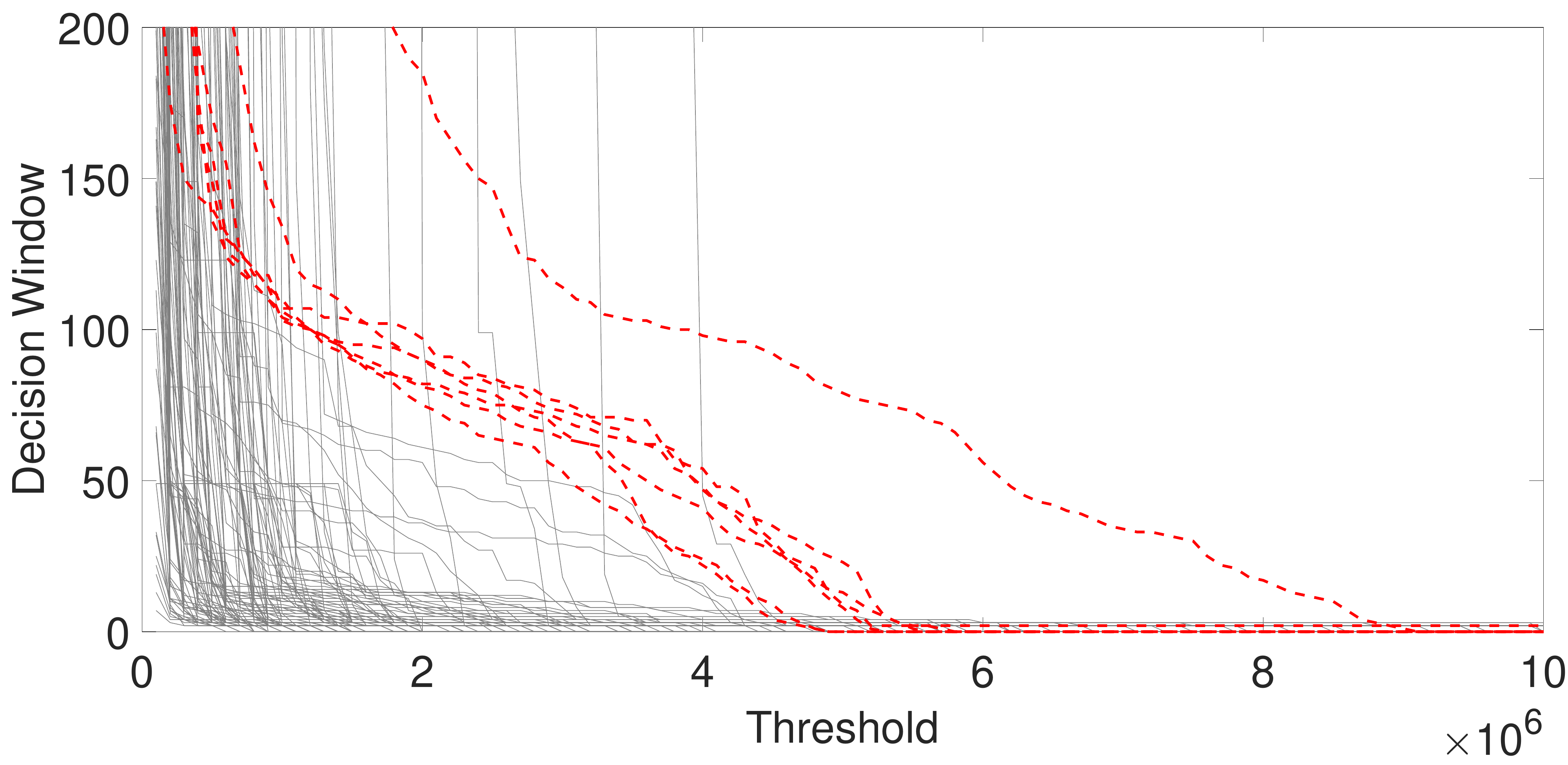}
	\caption{Threshold vs. Decision Window for benign applications (gray) and attack executions (red) using LSTM model in laptop scenario}
	\label{fig:laptop_lstm}
\end{figure}

\begin{figure}[!t]
	\centering
	\includegraphics[width=0.30\textwidth]{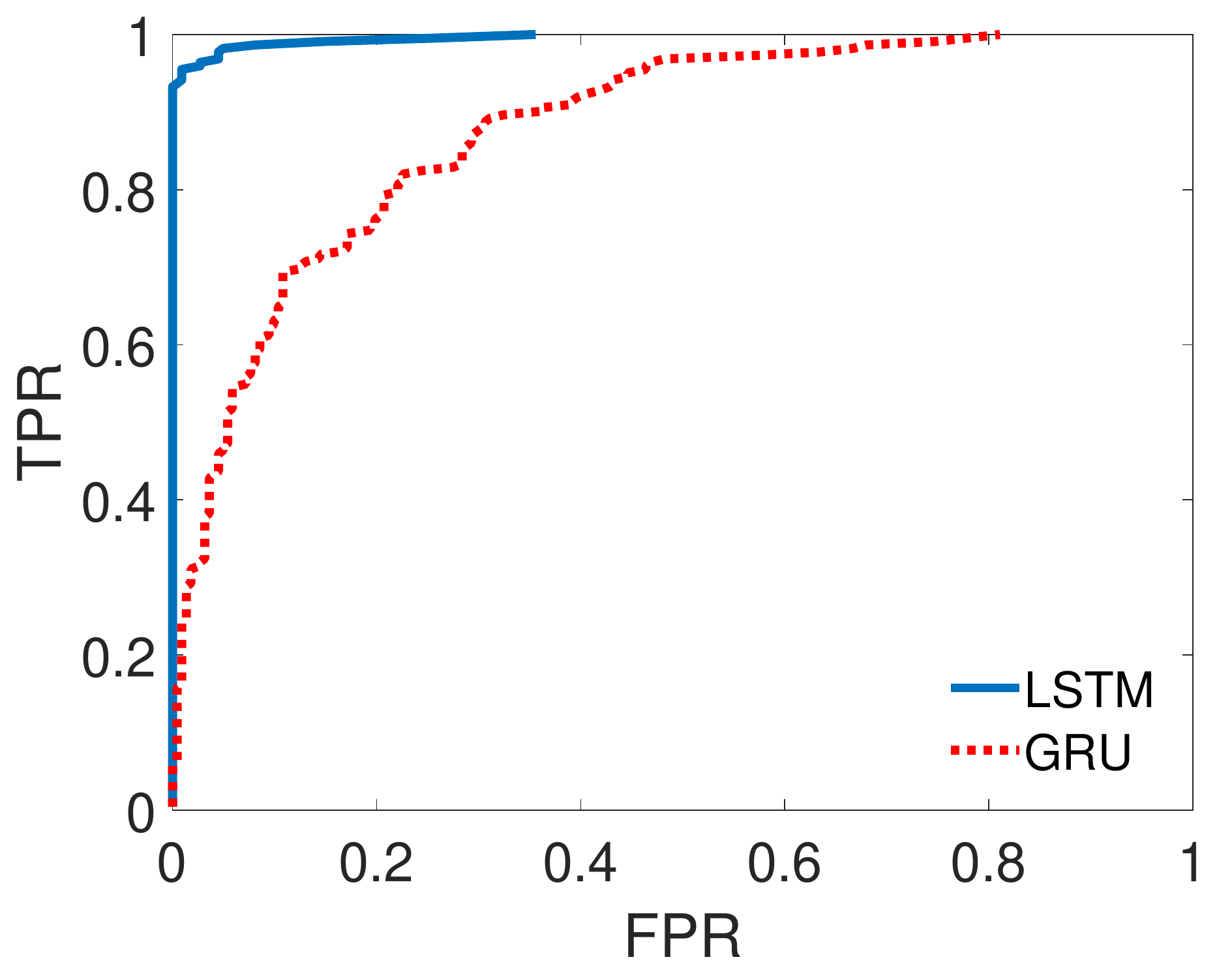}
	\caption{ROC curve for LSTM and GRU models in laptop scenario}
	\label{fig:laptop_ROC}
\end{figure}

When we analyze the relation between $D$ and $\tau_{A}$, we observe the same situation as in the server scenario. The lower $\tau_{A}$ values are not sufficient to differ the anomalies from benign executions. Therefore, we need to choose the optimal $\tau_{A}$ value slightly higher than the server scenario with a value of $3.8 \times 10^6$. The corresponding $D$ value is 60, which means that the anomalies are detected in 60 ms. The decision window is 10 ms bigger than server scenario however, the performance of \detector~is better in laptop scenario. In~\autoref{fig:laptop_ROC}, the ROC curves of LSTM and GRU models are compared. The AUC value of LSTM model is considerably higher than GRU model  with a value of 0.9865. However, the AUC value for GRU is 0.8925. This shows that LSTM outperforms GRU model to predict the counter values of benign applications. This also concludes that FNR and FPR are lower for LSTM models. 

Among the attack executions, Rowhammer attack can be detected with 100\% success rate since the prediction error is very high. The other attacks have similar prediction errors, hence, \detector~can detect the attacks with the same success rate. Since the computational power of laptop devices is low, the concurrent running applications have more noise on the counter values. Therefore, the prediction of the counter values is more difficult for RNN algorithms. While LSTM networks have small FPR for 4 and 5 applications running at the same time, GRU networks are not efficient to classify them as benign applications. 



The overall results show that LSTM works better than GRU networks for both laptop and server scenarios in~\autoref{table:accuracy}. The first and second values represent the LSTM and GRU false alarm rates per second in percentages, respectively. In the server scenario, videos, MySQL and Office applications never give false positives. Websites running in Google Chrome have a small amount of false alarm. Therefore, the FPR and FNR are around 0.12\% per second for LSTM network in server scenario overall. The main disadvantage of GRU networks is the poor performance in the prediction when the number of applications increases. The FPR and FNR are approximately 0.24\%. This shows that the number of false alarms is twice more for GRU based \detector.

In laptop scenario, LSTM performs better, which is supported by the false alarm rate. The number of false alarms is lower than the server scenario for laptop devices with a value of 0.09\%. On the other hand, the GRU networks are lack of ability to predict the counter values, thus, it is also reflected in false alarm rates. For every application, GRU has a higher false alarm rate than LSTM networks. Therefore, it is concluded that \detector should be trained with LSTM networks to have the better performance in both server and laptop scenarios.

\begin{table}[!t]
  \caption{The False Alarm Rate in percentage per second for applications}
  \begin{center}
    \begin{tabular}{|c|c|c|c|c|}
    \hline
      & \multicolumn{2}{|c|}{\bfseries {Server (\%)}} & \multicolumn{2}{|c|}{\bfseries{Laptop (\%)}}\\
      \hline
      & LSTM & GRU & LSTM & GRU\\
      \hline
      Benchmarks  & 0.1400&0.1442 & 0.1202&0.4808\\
      Websites  & 0.0550&0.0972 & 0.0278&0.6667\\
      Videos  & 0.0000&0.0000 & 0.0000&0.4138\\
      MySQL  & 0.0000&0.0000 & 0.0000&0.0000\\
      Apache  & 0.0000&0.8333 & 0.0000&0.4030\\
      Office  & 0.0000&0.0000 & 0.0000&0.0000\\
      2 Apps  & 0.0000&0.0000 & 0.0000&0.5000\\
      3 Apps  & 0.0000&0.1667 & 0.0000&0.5715\\
      4 Apps  & 0.0750&0.2000 & 0.0500&0.6667\\
      5 Apps  & 0.1250&0.2333 & 0.1000&0.8333\\
      \hline
    \end{tabular}
  \end{center}
  \label{table:accuracy}
\end{table}

\subsection{Varying size of Sliding Window}\label{sec:window_size}

We observed that the prediction results are affected by the size of the window. Therefore, we analyze the effect of sliding window size on anomaly detection with the data collected from core counters with 1 ms sampling rate in the server environment. 12 different window sizes are used to train LSTM and GRU models. The window size starts from 25 and increased by 25 at each step until reaching 300. 

\begin{figure}[!t]
	\centering
	\includegraphics[width=0.47\textwidth]{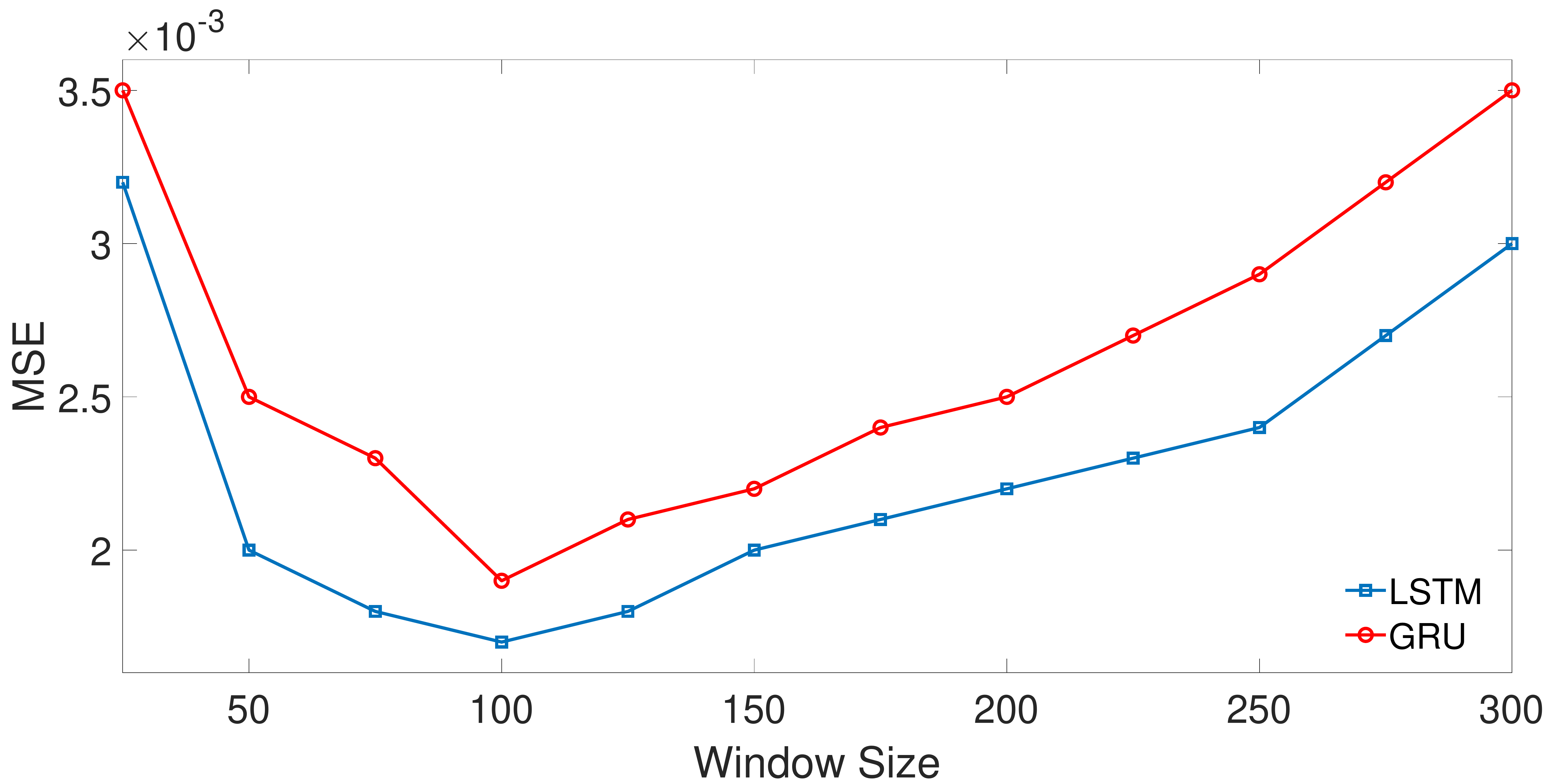}\label{fig:LSTM_GRU_error}
	\caption{The validation error with varying size of sliding window}
	\label{fig:lstm_gru_error}
\end{figure}

The changes in the validation error for both LSTM and GRU networks are depicted in~\autoref{fig:lstm_gru_error}. The overall GRU training error is higher than LSTM network for each window size. Both models reach the lowest error when the sliding window size is 100. Even though LSTM and GRU are designed to learn long sequences, it is recommended to choose the window size between 50-150. Since the best error is obtained with a window size of 100, all the models in the previous experiments are trained with this parameter. It is also important to note that the training time increases proportional to the size of the window.

\subsection{Time consumption for Testing}\label{sec:dynamic}

The dynamic detection of the anomalies also depends on the time consumption of predicting the next counter values. Therefore, the sampling rate should be chosen as close as to the timing consumption of predicting the next value. In our experiments, we observed that the prediction time is proportional to the size of the model. Since GRU has less number of cells in the architecture, the prediction of GRU is faster. While LSTM outputs prediction values for 3 counters in 2 ms, the prediction time for GRU is 1.7 ms. It shows that GRU is 15\% faster than LSTM in the prediction phase. However, due to the high FPR of GRU networks \detector~is trained with LSTM networks to detect anomalies in the real time system.

\subsection{Performance Overhead}\label{sec:performance_overhead}

The performance overhead of the proposed countermeasures is one of the most important concerns, since it affects all the applications running on the system. In this section, we evaluate the performance overhead for both server and laptop devices when core counters are used to collect data. The overhead amount is obtained with sampling rates of 1\,ms and 10\,$\mu$s. As it is expected the performance overhead increases in parallel with the sampling rate. In the server environment the overhead is around 7.7\% when the sampling rate is chosen as 10\,$\mu$s. The overhead of individual tests fluctuates between 1\% and 33\% for benign applications. The performance of system and memory benchmarks is affected more than processor based benchmarks. On the other hand, when the sampling rate is decreased to 1ms, the performance overhead drops to 3.5\%. The individual overheads change between 0.3\% and 18\%, which is more stable than the previous case.




In laptop scenario, the performance overhead is also calculated with the same benign benchmarks. The number of cores is smaller than the server scenario. Therefore, the performance monitoring unit only needs to read the counter values from 4 threads in parallel. On the other hand, since the system has lower features compared to the server machine, the overhead is increased when the sampling rate is increased .The overall performance degradation is 24.88\% for 10\,$\mu$s. The overhead fluctuates heavily, which means that the applications suffer from the frequent interruptions to read the counter values. Once the sampling rate is decreased to 1 ms, the overhead drops to 1.6\%, which is applicable in real-time systems. This overhead is also lower than the server scenario. Therefore, We preferred 1 ms sampling rate in our experiments.


\section{Comparison of \detector~with Prior Detection Methods}\label{sec:comparison}

There are several studies focused on microarchitectural attack detection as given in~\autoref{table:comparison}. While some works~\cite{briongos2018cacheshield,chiappetta2016real,Zhang2016cloudradar} use unsupervised techniques, Mushtaq et al.~\cite{mushtaq2018nights} benefits from supervised ML methods. All proposed methods claim that the false positive rate is very low in a real world scenarios. However, all these detection techniques are only applied for cryptographic implementations (AES, RSA, ECDSA etc.) and specific cache attacks (F+R, F+F, P+P). The performance of these techniques in real world scenarios (noisy environment, multiple concurrent processes) against transient execution attacks (Meltdown, Spectre, Zombieload etc.) and Rowhammer is questionable. In order to evaluate 4 proposed methods and \detector, we collected 6 million samples (4 million benign executions, 2 million attack executions) with 1ms sampling rate from 10 benign processes and 7 microarchitecture attacks by using \textbf{system-wide counters}. Note that each benign and attack execution is monitored 100 times in the server environment. The benign processes are chosen from diverse set of applications such as Apache, MySQL, browser and cryptographic implementations. The attacks cover cache-based, transient execution and Rowhammer attacks given in~\autoref{table:tests}. The detection algorithms from previous works are rewritten in Matlab environment and tested with the collected data.

\para{CPD from Briongos et al.~\cite{briongos2018cacheshield}} The first approach is Change Point Detection (CPD) which was implemented by Briongos et al.~\cite{briongos2018cacheshield} to detect the anomalies in the victim process. The primary advantage of the method is to have the capability of self-learning by observing the number of cache misses. On the other hand, the assumption of having almost no LLC miss is a strong assumption, which is not applicable in real-world scenarios for system-wide profiling. Especially, when an application runs for the first time in the system, the number of cache misses increases drastically. This yields to high number of false positives at the beginning of the applications. Even though it is tried to eliminate the initial false positives by increasing the initial value of cache misses under attack ($\mu_{a}$), we still observe several false positives at the beginning. It is also difficult to monitor each PID in the system since there are hundreds of processes running at the same time. 

For the evaluation of CPD method, we use the initial value of $\mu_{a} = 100$ and $\beta = 0.65$. When CPD method is applied to our dataset, we observe that the FPR is 3\% and FNR is 10\%. Therefore, the F-score is 0.9372. However, with the increasing number of concurrent processes, the false positive rate increases. This shows that CPD method is efficient for low system load however, it gives more false positives with increasing workload. The estimated detection time is around 300 ms for attack executions. The detection performance for Rowhammer and P+P attacks is poor since the number of cache misses is not high compared to benign processes. Therefore, these two attack types increase the FNR. 

\para{DTW from Zhang et al.~\cite{Zhang2016cloudradar}} The second detection method was proposed by Zhang et al.~\cite{Zhang2016cloudradar}, which benefits from Dynamic Time Warping (DTW) to detect the cryptographic implementations and then, the LLC hit and miss counters are monitored to detect the attacks. In the first step, DTW is used to compare the test data and the signature of cryptographic implementations obtained from branch instructions. Secondly, when the distance between test and target execution is very small, the LLC hit and LLC miss counters are monitored. If there is a sudden jump in these two counters, the anomaly flag is set. Again, this approach requires the PID of the monitored process.

In the evaluation of the method, we started with the application detection. Since the number of target applications is small in our dataset, DTW can detect them with 100\% success rate in a noiseless environment. However, when there is a concurrent process running in the system, the DTW distance is always high. The reason behind this failure is that branch instructions are heavily affected by the other processes. Therefore, DTW is not suitable for real-world scenarios. Another drawback is that if the microarchitectural attack already started, the branch instructions is also affected, which prevents to detect the target application. Hence, the attack detection step never starts. When the target process is detected, the anomaly detection step begins. If another concurrent work starts running at the same time, the cache miss and hit counters start increasing, which increases the FPR extremely. Since there is only a simple threshold approach to detect the attacks and the proposed decision window (5 ms) is too small, the FPR raises. In these circumstances, the approach achieves 10\% FPR. The attack detection is also not great since it is not possible to detect F+F attack with cache miss and hit counters. Thus, the the FNR increases in parallel which yields to 20\% FNR. Overall, the detection technique has 0.8572 F-score.

\para{PDF from Chiappetta et al.~\cite{chiappetta2016real}} In the third study, we evaluate the performance of normal distribution and probability density function, which is proposed by Chiappetta et al.~\cite{chiappetta2016real}. The detection technique monitors five counters (total instructions, CPU cycles, L2 hits, L3 miss and L3 hits) to catch the anomalies in the cryptographic implementations. This technique is used in unsupervised manner by only learning the normal distribution of the attack execution (F+R) with its mean and variance in the system. After the normal distribution is calculated, the probability density function (pdf) of both attack and benign executions is calculated for each counter sample. Then, an optimal threshold ($\epsilon$) is chosen to separate the benign and attack processes. To evaluate the performance of the method, we collected a separate dataset with the aforementioned five counters. The results indicated that total instructions and L2 hits decrease the performance of detecting anomalies. On the other hand, L3 hits and miss counters overperform other counters. The main drawback of this method is that there is no learning and the decisions are made on only cache miss and variance values. Therefore, when there is a benign application with high variance and mean, it is more likely to be classified as an anomaly. Especially, Apache server benchmark and videos running in browsers give high FPR. It is also observed that the P+P, F+F and Rowhammer attacks are not detected with a high accuracy, which give 0.2145 and 0.3732 for FPR and FNR, respectively. The F-score of the detection technique is 0.7278.

\para{OC-SVM from Mushtaq et al.~\cite{mushtaq2018nights}} The last method to compare is One Class Support Vector Machine (OC-SVM), which is used by Mushtaq et al.~\cite{mushtaq2018nights} to detect the anomalies on cryptographic implementations. The scope is limited F+F and F+R attacks. The number of counters tested in~\cite{mushtaq2018nights} is higher than three, which makes it impossible to monitor all of them concurrently. Therefore, we chose three counters (L1 miss, L3 hit and L3 total cache access), which give the highest F-score. Even though OC-SVM was used in a supervised way in~\cite{mushtaq2018nights}, we used it in unsupervised manner to maintain the consistency in the comparison. In the training phase, the model is trained with the 50\% of the benign execution data. Then, the attack and benign dataset are tested with the trained model. The obtained confidence scores are used to find the optimal decision boundary to separate the benign and attack executions. The optimal decision boundary shows that the FPR and FNR are 0.2750 and 0.2778, respectively. The main problem is that OC-SVM is not sufficient to learn the diverse benign applications, which increases the FPR drastically. Moreover, Rowhammer and F+F attacks are not detected, which is the reason of higher FNR. Therefore, the F-score remains at 0.7240.

\para{\detector} Finally, we apply \detector~to detect the anomalies in the system. Since the diversity of the benign executions is smaller in the comparison dataset, it is more easier to learn the patterns. It is also important to note that 50 measurements from each benign application is enough to reach the minimum prediction error. Once the LSTM model is trained with the benign applications, the attack executions and remaining benign application data are tested. The FPR and FNR remains at 0.2\% and 0.4\%, respectively. The F-score is 0.997 for the \detector.

\begin{table}[h!]
  \caption{Comparison of previous methods}
  \begin{center}
    \begin{tabular}{|c|c|c|}
    \hline
      & \textbf{Technique} & \textbf{F-score}\\
      \hline
      Briongos et al.~\cite{briongos2018cacheshield}  & CPD & 0.9372\\
      Zhang et al.~\cite{Zhang2016cloudradar}  & DTW & 0.8572\\
      Chiappetta et al.~\cite{chiappetta2016real}  & Normal Dist. & 0.7278\\
      Mushtaq et al.~\cite{mushtaq2018nights}  & OC-SVM & 0.7240\\
      \textbf{Our work}  & \textbf{LSTM/GRU} & \textbf{0.9970}\\
      \hline
    \end{tabular}
  \end{center}
  \label{tab:comp_tech}
\end{table}

The comparison results are summarized in~\autoref{tab:comp_tech}. The lack of appropriate learning is significant in the wild. It is also obvious that even simple learning algorithm such as CPD can help to overperform other detection techniques. We also show that the detection accuracy increases by learning the sequential patterns of benign applications with the system-wide profiling. Therefore, it is significantly important to extract the fine-grained information from the hardware counters to achieve the low FPR and FNR. The common deficiencies of previous works are listed below:

\begin{itemize}
    \item The detection methods focus on only cryptographic implementations, and the latest attacks such as Rowhammer, Spectre, Meltdown and Zombieload are not covered.
    \item There is no advanced learning technique applied in the detection methods. They mostly rely on the sudden changes in the counters, which increases the FPR heavily.
    \item The detection methods are either tested under no noise environment or the workload is not realistic. In addition, the FPR is not tested with a diverse set of applications.
\end{itemize}

\section{Discussion}\label{sec:discussion}

\para{Bypassing \detector} One of the questions about dynamic detection methods is that how an educated adversary can bypass the detection model? The common way is to put some delays between the attack steps to avoid increasing the counter values. For this purpose, we inserted different amounts of idle time frames between attack steps in Flush+Flush, Prime+Probe and Flush+Reload. We observed that the prediction errors in GRU and LSTM networks increases in parallel with the amount of sleep due to the high fluctuation. This shows that introducing delays between attack steps is not an efficient way to circumvent \detector. The reason behind this is the fluctuation in the time series data is not predicted well in the prediction phase. Therefore, we concluded that putting different amount of sleep between the attack steps is not enough to fool \detector. On the other hand, crafting adversarial examples is an efficient way to bypass Deep Learning based detection methods. For instance, Rosenberg et al~\cite{rosenberg2017generic} shows that LSTM/GRU based malware detection techniques can be bypassed by carefully inserting additional API calls in between. Therefore, crafting adversarial code snippets to change the performance counters in the attack code may fool \detector. The main difficulty in this approach is that it is not possible to decrease the counter values by executing more instructions between attack steps. Therefore, applying adversarial examples on hardware counter values is not trivial.

\para{Training Algorithm} \detector~investigates both available long-term dependency learning techniques. We observed that GRU performs worse than LSTM networks to predict the counter values in the next time steps. This is because of the lack of internal memory state, which keeps the relevant information from previous cells. This result is also supported with the high FPR and FNR of GRU networks. Since the prediction error increases for attack executions more than benign applications, the detection accuracy decreases. Therefore, we recommend to train LSTM networks for microarchitectural attack detection techniques. 

\para{Dynamic Detection} The current implementation requires to have a GPU to train \detector, as GPU based training 40 times faster than CPU based training. The training is mostly done in an offline phase and it does not affect the dynamic detection. On the other hand, dynamic detection heavily depends on the matrix multiplication, since the trained model is loaded as a matrix in the system and the same matrix is multiplied with the current counter values. Hence, the required time to predict the next counter values is lower. In addition, we observed that the performance overhead is negligible for the matrix multiplication in the CPU systems. Therefore, \detector\ can be implemented in server/cloud/laptop environments, even though there is no GPU integrated in the system.

\section{Conclusion}\label{sec:conclusion}

This study presented \detector, which exploits the power of neural networks to overcome the limitations of the prior works, and further proposes a novel generic model to classify microarchitectural events. \detector~is able to dynamically detect microarchitectural anomalies in the system through learning benign workload. In our study, we adopted two state-of-the-art RNN models: GRU and LSTM. We concluded that LSTM is more preferable compared to GRU for our use case. Further, the number of measurements and the sliding window size have a significant effect on the validation error in training phase, which makes it crucial to choose the optimal values to have better prediction results. \detector~is applicable to both server and laptop environments with a high accuracy. In order to evaluate the performance of \detector, we used both benchmarks and real-world applications and achieved 0.12\% and 0.09\% FPRs for server and laptop environments, respectively. \detector~is also tested against previous works in the realistic scenarios and it is concluded that, \detector overperforms other detection meahcnisms in the wild. While the performance overhead in laptop environment is less than server, \detector~is still applicable in the real world systems with minimal overhead.

%
%
\section*{Acknowledgments}

This work is supported by the National Science Foundation, under grants CNS-1618837 and CNS-1314770.

%
%

{\footnotesize \bibliographystyle{plain}
\bibliography{main}}



%
\newpage

\section{Appendix}
\label{sec:appendix}

\subsection{Tables for Performance Counters and Benchmarks}

\begin{table}[h]
    \scriptsize
	\caption{Counter Selection for core counters}
	\centering
	\begin{tabular}{| c | c |}
		\hline
		Counter & F-score\\
		\hline
		$Dtlb\_Load\_Misses.Miss\_Causes\_A\_Walk$ & 0.5657 \\
		$Dtlb\_Load\_Misses.Walk\_Completed\_4K$ & 0.5226 \\
		$Dtlb\_Load\_Misses.Walk\_Completed$ & 0.5327 \\
		$Dtlb\_Load\_Misses.Walk\_STLB\_Hit\_4K$ & 0.3627\\
		$UOPS\_Issued\_Any$ & 0.3663\\
		$ICACHE.Hit$ & \textbf{0.8137}\\
		$ICACHE.Miss$ & \textbf{0.7979}\\
		$L1D\_Pend\_Miss.Pending$ & 0.6818\\
		$L1D\_Pend\_Miss.Request\_FB\_Full$ & 0.6698\\
		$L1D.Replacement$ & 0.7523\\
		$L2\_Rqsts\_Lat\_Cache.Miss$ & 0.6244 \\
		$LLC\_Miss$ & \textbf{0.8416}\\
		$LLC\_Reference$ & 0.6167\\
		$IDQ.Mite\_UOPS$ & 0.3383\\
		$BR\_Inst\_Exec.Nontaken\_Cond.$ & 0.2703 \\
		$BR\_Inst\_Exec.Taken\_Cond.$ & 0.3390 \\
		$BR\_Inst\_Exec.Taken\_Direct\_Jmp$ & 0.3455\\
		$BR\_Inst\_Exec.Taken\_Indirect\_Jmp\_Non\_Call\_Ret$ & 0.3137  \\
		$BR\_Inst\_Exec.Taken\_Indirect\_Near\_Return$ & 0.2944  \\
		$BR\_Inst\_Exec.Taken\_Direct\_Near\_Call$ & 0.3618 \\
		$BR\_Inst\_Exec.Taken\_Indirect\_Near\_Call$ & 0.3592 \\
		$BR\_Inst\_Exec.All\_Cond.$ & 0.2634\\
		$BR\_Inst\_Exec.All\_Direct\_Jmp$ & 0.3238 \\
		$BR\_Misp\_Exec.Nontaken\_Cond.$ & 0.3648 \\
		$BR\_Misp\_Exec.Taken\_Cond.$ & 0.4510 \\
		$BR\_Misp\_Exec.Taken\_Indirect\_Jmp\_Non\_Call\_Ret$ & 0.4455  \\
		$BR\_Misp\_Exec.Taken\_Ret\_Near$ & 0.3491  \\
		$BR\_Misp\_Exec.Taken\_Indirect\_Near\_Call$ & 0.3553 \\
		$BR\_Misp\_Exec.All\_Branches$ & 0.2700 \\
		$BR\_Inst\_Retired.Cond.$ & 0.4623 \\
		$BR\_Inst\_Retired.Not\_Taken$ & 0.4412 \\
		$BR\_Inst\_Retired.Far\_Branch$ & 0.4608  \\
		$BR\_Misp\_Retired.All\_Branch$ & 0.4615  \\
		$BR\_Misp\_Retired.Cond.$ & 0.3786 \\
		$BR\_Misp\_Retired.All\_Branches\_Pebs$ & 0.2111  \\
		$BR\_Misp\_Retired.Near\_Taken$ & 0.2871 \\\hline
	\end{tabular}
	\label{table:counters}
\end{table}


\begin{table*}[t]
	\caption{Benchmark tests used in the experiments}
	\centering
	\scriptsize
	\tabcolsep=0.11cm
	\begin{tabular}{| l | l | l | l | l | l | l | l |}
		\hline
		\multicolumn{3}{|c|}{\bfseries Processor Tests} & \bfseries System Tests & \bfseries Disk Tests & \bfseries Memory Tests & \bfseries Real-World & \bfseries Attacks\\
		\hline
		1) Aobench&          41) Minion 1         & 81) Graphics 1  & 120) Apache        & 153) Aio-stress   & 165) Mbw    & 174) Websites   & 1) Flush+Flush\\
		2) Botan 1&          42) Minion 2         & 82) Graphics 2  & 121) Battery & 154) Blogbench 1  & 166) Ram 1 & 175) Videos & 2) Flush+Reload\\
		3) Botan 2&          43) Minion 3         & 83) Graphics 3  & 122) Compress & 155) Blogbench 2  & 167) Ram 2 & 176) MySQL  & 3) Prime+Probe\\
		4) Botan 3&          44) Perl 1 & 84) Graphics 4  & 123) Git           & 156) Compile & 168) Ram 3 & 177) Apache & 4) Meltdown\\
		5) Botan 4&          45) Perl 2 & 85) Graphics 5  & 124) Hint          & 157) Dbench       & 169) Ram 4 & 178) Office & 5) Spectre\\
		6) Botan 5&          46) Radiance 1       & 86) Graphics 6  & 125) Nginx         & 158) Fio 1        & 170) Ram 5 & & 6) Rowhammer\\
		7) Bullet 1&         47) Radiance 2       & 87) Graphics 7  & 126) Optcarrot     & 159) Fio 2        & 171) Stream & & 7) Zombieload\\    
		8) Bullet 2&         48) Scimark 1        & 88) Hpcg               & 127) Php 1         & 160) Iozone       & 172) T-test & & \\     
		9) Bullet 3&         49) Scimark 2        & 89) Luajit 1           & 128) Php 2         & 161) Postmark     & 173) Tinymem & &\\
		10) Bullet 4&        50) Scimark 3        & 90) Luajit 2           & 129) Pybench       & 162) Sqlite    & &   &\\                    
		11) Bullet 5&        51) Scimark 4        & 91) Luajit 3           & 130) Schbench      & 163) Tiobench  & &   &\\                    
		12) Bullet 6&        52) Scimark 5        & 92) Luajit 4           & 131) Stress-ng 1   & 164) Unpack  & & &\\                    
		13) Bullet 7&        53) Scimark 6        & 93) Luajit 5           & 132) Stress-ng 2 & & & &\\   
		14) Cache 1&    54) Swet             & 94) Luajit 6           & 133) Stress-ng 3 & & & &\\   
		15) Cache 2&    55) Hackbench        & 95) Mencoder           & 134) Stress-ng 4 & & & &\\   
		16) Cache 3&    56) M-queens         & 96) Multichase 1       & 135) Stress-ng 5 & & & &\\   
		17) Gzip&   57) Mrbayes          & 97) Multichase 2       & 136) Stress-ng 6 & & & &\\   
		18) Dcraw&           58) Npb 1            & 98) Multichase 3       & 137) Stress-ng 7 & & & &\\   
		19) Encode&      59) Npb 2            & 99) Multichase 4       & 138) Stress-ng 8 & & & &\\   
		20) Ffmpeg&          60) Npb 3            & 100) Multichase 5      & 139) Stress-ng 9 & & & &\\   
		21) Fhourstones&     61) Npb 4            & 101) Polybench-c       & 140) Stress-ng 10 & & & &\\  
		22) Glibc 1&   62) Npb 5            & 102) Sample    & 141) Stress-ng 11 & & & &\\ 
		23) Glibc 2&   63) Npb 6            & 103) Sudokut           & 142) Stress-ng 12  & & & &\\  
		24) Glibc 3&   64) Npb 7            & 104) C-ray             & 143) Stress-ng 13 & & & &\\  
		25) Glibc 4&   65) Povray           & 105) Cloverleaf        & 144) Stress-ng 14 & & & &\\  
		26) Glibc 5&   66) Smallpt          & 106) Dacapo 1     & 145) Stress-ng 15 & & & &\\  
		27) Glibc 6&   67) Tachyon          & 107) Dacapo 2     & 146) Stress-ng 16 & & & &\\  
		28) Glibc 7&   68) Bork             & 108) Dacapo 3     & 147) Sunflow & & & &\\       
		29) Glibc 8&   69) Build Apache     & 109) Dacapo 4     & 148) Sysbench 1 & & & &\\    
		30) Gnupg&           70) Byte 1           & 110) Dacapobench 5     & 149) Sysbench 2 & & & &\\    
		31) Java 1& 71) Byte 2           & 111) John 1 & 150) Tensorflow & & & &\\    
		32) Java 2& 72) Byte 3           & 112) John 2 & 151) Tjbench & & & &\\       
		33) Java 3& 73) Byte 4           & 113) John 3 & 152) Xsbench & & & &\\       
		34) Java 4& 74) Clomp            & 114) Mafft & & & &  &\\             
		35) Java 5& 75) Crafty           & 115) N-queens & & & & &\\          
		36) Java 6& 76) Dolfyn           & 116) Openssl & & & & &\\           
		37) Lzbenc 1&       77) Espeak           & 117) Primesieve & & & & &\\        
		38) Lzbench 2&       78) Fftw             & 118) Stockfish & & & & &\\         
		39) Lzbench 3&       79) Gcrypt           & 119) Ttsiod & & & & &\\   
		40) Lzbench 4&       80) Gmpbench & & & & & &\\
		\hline
	\end{tabular}
	\label{table:tests}
\end{table*}

\end{document}